\documentclass{emulateapj}
\usepackage{graphics}    

\newcommand{\Ngals}{N$_{\rm gals}$}
\newcommand{\Mstar}{$M_{*}$}
\newcommand{\Ndense}{${\cal N}$}
\newcommand{\rdelta}{$R_{\Delta}$}
\newcommand{\rN}{$R_{200}^{\cal N}$}
\newcommand{\rtwo}{$R_{200}$}
\newcommand{\mypdf}{PDF(\Ngals, $r$, $M$)}
\newcommand{\mycpdf}{PDF$_{C}$(\Ngals, $r$, $M$)}
\newcommand{\cg}{c$_{gal}$}
\newcommand{\cdm}{c$_{DM}$}
\newcommand{\DNcrit}{$\Delta$\Ndense$_{crit}$}
\newcommand{\DNmean}{$\Delta$\Ndense$_{mean}$}
\newcommand{\twoOM}{200$\Omega_{m}^{-1}$}
\newcommand{\Lmax}{$\cal L_{\rm max}$}
\newcommand{\Lz}{$\cal L$$(z)$}
\newcommand{\Lbcg}{$\cal L_{\rm BCG}$}
\newcommand{\z}{$z$}
\newcommand{\griz}{$g$, $r$, $i$, and $z$}

\slugcomment{Accepted for publication in the Astrophysical Journal July 3 2005}
\shorttitle{R$_{200}$, Radial Profiles and Luminosity Functions for SDSS Galaxy Clusters}
\shortauthors{Hansen et al.}

\begin{document}

\title{Measurement of Galaxy Cluster Sizes, Radial Profiles, and Luminosity Functions from SDSS Photometric Data}

\author{Sarah M. Hansen\altaffilmark{1,2}, 
Timothy A. McKay\altaffilmark{3}, 
Risa H. Wechsler\altaffilmark{1,2,4}, \\ 
James Annis\altaffilmark{5}, 
Erin Scott Sheldon\altaffilmark{2}, and
Amy Kimball\altaffilmark{6}}
\altaffiltext{1}{Department of Astronomy and Astrophysics,
University of Chicago, Chicago, IL 60637} 
\altaffiltext{2}{Kavli Institute for Cosmological Physics,
University of Chicago, Chicago, IL 60637} 
\altaffiltext{3}{Physics Department, University of Michigan, Ann
  Arbor, MI 48109}
\altaffiltext{4}{Hubble Fellow}
\altaffiltext{5}{Fermi National Accelerator Laboratory, Batavia, IL 60510}
\altaffiltext{6}{Astronomy Department, University of Washington,
  Seattle, WA 98195}

\begin{abstract}
  
  Imaging data from the Sloan Digital Sky Survey is used to measure
  the empirical size-richness relation for a large sample of galaxy
  clusters.  Using population subtraction methods, we determine the
  radius at which the cluster galaxy number density is \twoOM\ times
  the mean galaxy density, without assuming a model for the radial
  distribution of galaxies in clusters. If these galaxies are unbiased
  on Mpc scales, this galaxy-density-based \rtwo\ reflects the
  characteristic radii of clusters. We measure the scaling of this
  characteristic radius with richness over an order of magnitude in
  cluster richness, from rich clusters to poor groups.  We use this
  information to examine the radial profiles of galaxies in clusters
  as a function of cluster richness, finding that the concentration of
  the galaxy distribution decreases with richness and is
  systematically lower than the concentrations measured for dark
  matter profiles in N-body simulations.  Using these scaled radii, we
  investigate the behavior of the cluster luminosity function, and
  find that it is well matched by a Schechter function for galaxies
  brighter than $M_r=-18$ only after the central galaxy has been
  removed.  We find that the luminosity function varies with richness
  and with distance from the cluster center, underscoring the
  importance of using an aperture that scales with cluster mass to
  compare physically equivalent regions of these different systems. We
  note that the lowest richness systems in our catalog have properties
  consistent with those expected of the earliest-forming halos; our
  cluster-finding algorithm, in addition to reliably finding clusters,
  may be efficient at finding fossil groups.

\end{abstract}

\section{Introduction}

Galaxy groups and clusters appear in many guises. Observationally,
they can be identified as pools of X-ray emitting gas, collections of
galaxies, Sunyaev-Zel'dovich decrements of the CMB, or strong features
in the gravitational shear field.  Theoretically, they are identified
as the largest over-dense `halos' of dark matter. Mass is the key
defining attribute of a galaxy cluster. The evolution of the cluster
mass function and its variance plays an important role in constraining
cosmological parameters describing dark energy, such as $w$\ and
$\Omega_{\Lambda}$\ \citep[e.g.][]{Lim05}, and large-scale structure,
such as $\Omega_m$\ and $\sigma_8$ \citep[e.g.][]{Bah03b}. Mass
estimates and studies of cluster members both rely upon knowing the
size of a cluster of given mass. To define the mass of a cluster,
typically a cluster radius is specified through some prescription, and
the total mass taken to be the mass contained within that radius.
Various operational definitions are used to determine cluster size and
mass, but all of these show a regular increase in cluster size with
mass. In order to study objects in a wide range of masses in a
meaningful way, we must be able to determine an appropriate size scale
for these clusters and groups; we may then use this characteristic
scale as an aperture within which to make comparisons.

In numerical simulations, the precision of cluster mass and size
measurements is limited only by resolution.  Still, there are a
variety of definitions for both size and mass in use, as discussed in
some detail by \citet{Whi01}. One class of estimates is based on
top-hat filtered spherical over-densities. In this model, clusters are
expected to be virialized within regions where the enclosed mean mass
density exceeds the critical density by a factor $\Delta \sim$ 200
\citep{Pee93, Pea99}. The radius at which this over-density is
reached, $R_{\Delta=200}$, is used as the characteristic radius of the
cluster. The total mass within this radius, $M_{\Delta=200}$, is used
as the characteristic mass. A number of choices of $\Delta$ are in use
in the literature, from an overdensity of 180 times the mean
background \citep[e.g.][]{Jen01,Kra04}, 200 times the critical
density, i.e. 200$\Omega_{M}^{-1}$ times the mean background
\citep[e.g.][]{Dia01, Evr02, Koc03}, to the ``virial mass''
\citep{ECF96, Bul01}. Alternative definitions identify halos by
`friends-of-friends' (FOF) methods \citep{Dav85, Jen01}. In these
methods, particles are associated with halos to which they are linked
by sequences of neighboring particles. Masses for FOF halos are often
given by the sum of member particle masses, but as the halos are not
required to be spherical, halo size is less clearly defined.

Observationally, cluster mass and size are difficult to measure
directly; typically some mass estimator is adopted as a proxy, and
typically a mass model is assumed to calculate the virial size. With
deep observations, lensing may be used to make detailed mass maps of
rich clusters, but this is not yet a practical technique for a large
sample of systems spanning a wide range of masses. With large,
shallower surveys, the lensing signals from a set of lens systems can
be stacked to determine a composite mass profile \citep{She01}.
However, the large number of lens systems needed for high signal to
noise makes such a measurement difficult for examining narrow mass
ranges.  While examining the lensing signal from an individual cluster
recovers a specific mass, that estimate is affected by the errors,
such as those due to projection effects, in modeling the mass.
Combining the lensing signal from many clusters is advantageous
because projection effects are unimportant, and the stacking
simplifies modeling of neighboring structures. While stacking does
limit the resolution in mass for a large set of clusters, this
limitation is due to the amount of data available rather than the
technique used. Other observational techniques for estimating mass,
whether using X-ray temperature, SZ flux, or velocity dispersion as a
mass estimator, also rely upon either expensive spectroscopy to gain a
detailed understanding of a few rich systems, or upon models for the
mass distribution used to infer the total system mass and size. To
compare observational data to theoretical models, and to compare the
observed properties of clusters of different masses, it would be
preferable to avoid using a model-dependent mass/radius scaling.

With a large photometric optical survey it is now possible to use
cluster richness to characterize galaxy systems without suffering from
the projection effects that have plagued such a mass estimator in the
past. Since the cluster catalog of \citet{Abe58}, systems of galaxies
have been sorted and compared using a variety of richness parameters,
many of which have been based on the number of galaxies within a
certain luminosity range and distance from the estimated cluster
center.  The richness parameters of \citet{Abe89, Cou91, Dal92, Lum92,
  Lid96, Pos96, Ost98, Ols99, Gla02, Gotcat, Pos02, Gal03} are of this
type; see \cite{Bah81, Yee99, Bah03} for further discussion and
comparison of some of these richness estimators. We can stack systems
in narrow richness bins, and measure directly the distribution of
galaxies in clusters over a wide range of masses. This galaxy
distribution is used to estimate the virial size of these systems.
Ideally, cluster members would be identified spectroscopically.  Such
data is not feasible to obtain for a very large sample of clusters, so
we rely upon projected photometric data taken in multiple bandpasses,
and correct for the foreground and background galaxies that
contaminate our line of sight to each cluster.

The Sloan Digital Sky Survey \citep[SDSS,][]{Yor00,Sto02} data offer
thousands of clusters and groups for study, and can be used to measure
excellent photometric redshifts for those objects.  We use data from
the SDSS to directly determine a size-richness relation for groups and
clusters with a model-independent method. Since we cannot directly
measure the radius at which the cluster has a mass over-density of
${\Delta M}$, we instead determine the radius $R_{\Delta}^{\cal N}$ at
which the space density of cluster galaxies, \Ndense, is over-dense by
${\Delta}$\Ndense. We present the scaling of \rN\ with richness, which
can be employed to further study galaxy clusters and their members.
For example, to study the relationship between different mass
estimators, such as the mass from lensing measurements and the total
luminosity of the cluster, it is essential to know this scaling of
cluster size with richness.

If the distribution of galaxies in a halo traces the overall
dark matter distribution, our galaxy-density based \rN\ will reflect
the characteristic radius of clusters.  In detail, the relation
between the dark matter density profile and the radial distribution of
a population of galaxies depends on a number of physical processes
including dynamical friction and tidal striping, and depends on the
properties of the galaxy sample, but both simulations and previous
observational work suggest that the distribution of galaxies in a halo
at least roughly traces the overall dark matter distribution 
\citep[e.g.,][]{NK04, lin04}.

This hypothesis is supported by recent lensing and galaxy clustering
measurements \citep{she04,Wei04}, which suggest that the bias of typical
SDSS galaxies is approximately one and is roughly scale-independent on
scales larger than a few hundred $h^{-1}$ kpc.  In any case, without
assuming a model for the radial distribution of galaxies in clusters,
we can directly measure the radius at which the galaxy density in
clusters is $\Delta$ times more dense than the average background.

Our investigation does not distinguish between galaxy groups and
clusters.  Systems of galaxies come in a range of masses from single
galaxies in larger halos up to the most massive of clusters.  There is
a clear boundary between galaxies and systems of galaxies; there is
not any clear dividing point between poor and rich systems of galaxies
as observed in the optical. We use the full range of our cluster
finder to develop our catalog of systems; our mass estimator spans the
range from highly populated clusters down to very sparse systems, and
we refer to any system with two or more galaxies as a ``cluster''.
There is no strong break in the properties of the systems found as a
function of richness.

In \S \ref{data} we describe the SDSS data used and discuss the
cluster finding technique and richness measurement; in \S \ref{method}
we present and test our method of background subtraction through
examination of the radial profile and luminosity function. Our
calculation of \rN\ as a function of richness is presented in \S
\ref{r200}. We further examine the radial density profile within \rN\ 
in \S \ref{rpr200}; we examine the luminosity function within
\rN\ and how the luminosity function changes as a function of $r$/\rN\ for clusters of
different richnesses in \S \ref{lfr200}. Throughout, we assume a flat,
LCDM cosmology with H$_{0} = 100h$ km s$^{-1}$ Mpc$^{-1}$, $h =
0.7$, and matter density $\Omega_m = 0.3$.

\section{Data}\label{data}
\subsection{SDSS Galaxies}
In this study, we use 395 deg$^{2}$ of SDSS commissioning data
\citep{Yor00}, a subset of the Early Data Release data. In particular,
we use the 170 contiguous square degrees imaged September 19 and 25
1998, covering the range 145.1\degr\ $\leq$ RA $\leq$ 236.0\degr,
-1.25\degr\ $\leq$ DEC $\leq$ +1.25\degr\ (J2000), known as SDSS
stripe 10, and the 225 contiguous square degrees imaged March 20-21
1999, covering the range 351\degr\ $\leq$ RA $\leq$ 56\degr,
-1.25\degr\ $\leq$ DEC $\leq$ +1.25\degr\ (J2000), known as stripe
82. Seeing varies on these two stripes from 1.0" to 2.0", and the data
are photometrically uniform to within 3\% \citep{Hog01,
  Smi02}. Star-galaxy separation is robust to 21.0, 21.0, 21.0, and
19.8 in \griz\ passbands respectively \citep{Scr01}, which we
adopt as the limiting apparent magnitudes for this work. All apparent
magnitudes are measured by the photometric data processing pipeline
using a modified version of the \citet{Pet76} system (see
\citealt{Bla01} for a discussion of the advantages of Petrosian
magnitudes), and are corrected for Galactic extinction using the dust maps of
\citet{Sch98}. Further details about the photometric data and the
parameters measured may be found in \citet{Lup01, Lup05} and
\citet{Sto02}.

\subsection{Cluster Finding Technique} \label{maxBCG}

Clusters used in this study are detected by the maxBCG algorithm.
This method relies on the observation that clusters host a population
of early-type galaxies that have small dispersion in color
\citep{Bow92,Sta95,Sma98,Gla00}. These cluster members populate the
red sequence on a color-magnitude diagram. The brightest cluster
galaxies (BCGs) have colors that are compatible with the red sequence
galaxies \citep{Ara98, Nel02}, and also have a very small dispersion
in luminosity \citep{San02, San76,Oem76,Hoe80,Sch83,Pos95,Col98}. There is evidence that the colors
and magnitudes of BCGs remain predictable (and thus indicative of
redshift) until at least \z\ = 0.6 \citep{Ara98,Nel02}, shifting only
due to passive evolution.

The maxBCG algorithm takes advantage of these observed properties to
find BCGs and the galaxies that are associated with them. For every
galaxy, the algorithm calculates the highest likelihood, \Lmax, that
any given galaxy is a BCG (which means the galaxy has the properties
of a BCG plus has a red sequence around it), and then identifies
clusters by finding galaxies with high \Lmax\ values compared to the
surroundings. \Lmax\ for each galaxy is determined by finding the
maximum of \Lz, the calculated likelihood as a function of redshift
for the candidate. The likelihood function \Lz\ is the sum of two
terms, and is calculated at every redshift from \z\ = 0.0 to \z\ = 0.6
in steps of 0.01. That is, for each galaxy we calculate

\begin{equation}
\cal L_{\rm max} = \rm max\ \cal L(\rm z);\ \rm where\ 
\cal L(\rm z) = \cal L_{\rm BCG} + \rm \log N_{gals}.
\end{equation}

\begin{figure}
  \epsscale{1.1}\plotone{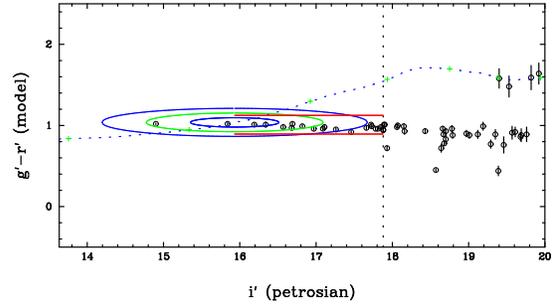} 
\caption[f1_color.eps]{Color-magnitude diagram of
    observed $g$ - $r$ versus apparent $i$-band for galaxies near a rich
    cluster at $z$ = 0.15. Ellipses represent 1-, 2-, and 3-$\sigma$
    contours around the mean BCG color and magnitude at that
    redshift. The dotted line indicates the track of BCG color and
    magnitude as a function of redshift. The horizontal lines and
    vertical dashed line show the region of inclusion for \Ngals\
    determination.
    \label{maxbcg}}
\end{figure}

The first term, \Lbcg, is the BCG likelihood: the likelihood that the
galaxy in question is consistent with, compared to the known
population dispersions, the apparent magnitudes and colors of the
mean BCG population as seen at that redshift. These colors and
magnitudes were derived from the properties of Abell clusters observed
by the SDSS.  The second term, $\log$\ \Ngals, is the log of the count
of the other galaxies in the vicinity that are also of the right color
and magnitude to be cluster members.  To count the number of galaxies,
we examine only those projected within 1$h^{-1}$ Mpc (at the redshift
in question) that fall in the red sequence for each redshift. A galaxy
is within the red sequence if it is within 2$\sigma$ ($\sigma = 0.05$
mags) of the mean BCG color at that redshift, fainter than the BCG
candidate, and brighter than $M_{i}$ = -20.25 (approximately 0.5
L$_{*}$). Figure \ref{maxbcg} shows a color-magnitude diagram for a
rich cluster. The region of inclusion for \Ngals\ determination shown 
by the contours; the color and apparent luminosity of a passively-evolving BCG
as a function of redshift is indicated by a dotted track. The
interplay between finding a galaxy with the right color and luminosity
to be a BCG and a galaxy with red sequence neighbors determines the
redshift distribution of \Lz. We find the redshift that maximizes \Lz\ 
for the galaxy in question. That galaxy is assigned \Lmax\ and the
corresponding redshift.  At the end of this process each galaxy in the
catalog has a single maximum likelihood and a single redshift.

We then need to select the cluster centers out of the catalog of all
galaxies.  We find the peaks in the distribution of \Lmax\ over all
galaxies; these peaks are the clusters. For each candidate BCG, we
check whether its likelihood \Lmax\ is the highest likelihood when
compared with neighboring candidates within $\Delta z = 0.05$ that are
projected within 1$h^{-1}$ Mpc of the BCG candidate in question. If
the candidate BCG's \Lmax\ is the greatest of those neighbors, we list
the candidate in the cluster catalog as a BCG.

\begin{figure*}
  \plotone{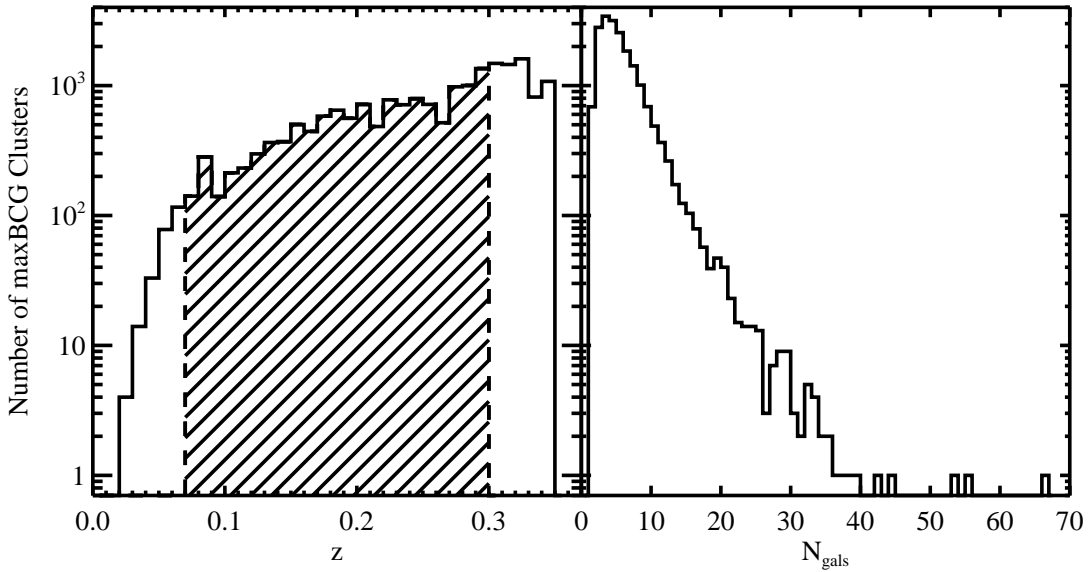} 
\caption[f2.eps]{ {\bf Left:} the redshift
    distribution of maxBCG-identified objects. We use clusters in the
    shaded region; simulations show that the completeness rate begins to
    drop for $z > 0.3$.  {\bf Right:} the distribution of $0.07 \le z
    \le 0.3$ maxBCG-identified objects as a function of cluster
    richness, \Ngals. There are 12,830 systems identified in this redshift range; 2270 of them have \Ngals\ $\ge 8$, and 19 clusters have \Ngals\ $\ge 30$. \label{distrib}}
\end{figure*}

The cluster catalog produced contains information about each
cluster identified, including the photometric properties of
the BCG, the estimated redshift and the richness, \Ngals. Each cluster center is taken to be at the location
of its BCG. The richness measurement for the cluster, \Ngals, is
defined to be the number of galaxies in the red sequence at the
derived redshift. The
resulting catalog contains objects over a wide range of richnesses:
from quite poor systems of only a few galaxies (\Ngals\ $\leq$ 8;
$\sigma_{v}$ $\le$ 300 km s$^{-1}$; 10,560 systems in 0.07 $\leq z \leq$
0.3) to very massive clusters of
hundreds of galaxies (\Ngals\ $\geq$ 30; $\sigma_{v} \ge$ 700 km
s$^{-1}$; 19 systems in 0.07 $\leq z \leq$ 0.3). Figure \ref{distrib} shows the distribution of identified
objects as a function of redshift and of richness. So as to avoid making an arbitrary distinction between a group and a cluster, we will generically refer to a system of galaxies as a cluster, and specify the value of \Ngals\ of that system.

The maxBCG algorithm has been tested extensively for completeness and
purity. All previously known Abell and NORAS X-ray clusters in the
region surveyed are recovered.  Simulations suggest that maxBCG
recovers and correctly estimates the richness for greater than 90\% of
clusters and groups present with \Ngals\ $\geq$ 15 out to a redshift
of \z\ = 0.3. The completeness and selection function of the algorithm will
be further explored in \citet{Wec05}. The clusters identified by this algorithm have been compared with the
objects found by different cluster-finding algorithms run on the same
dataset. Discussion of the differences between maxBCG and other
algorithms can be found in \citet{Bah03}.

\begin{figure}
  \plotone{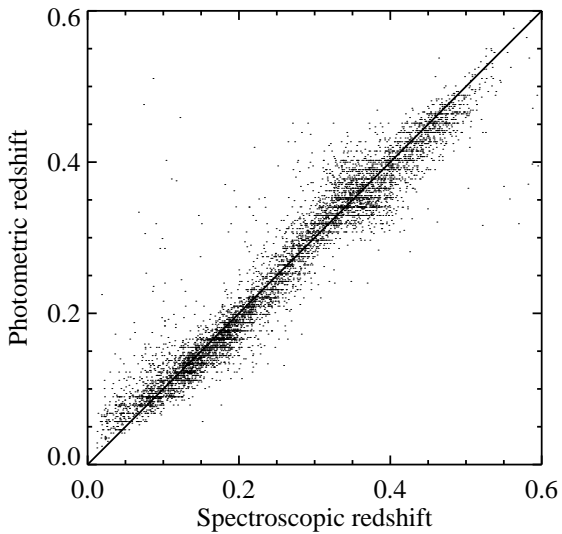} 
\caption[f3.eps]{Photometric redshift estimation
    for the maxBCG cluster finding algorithm is tested here by
    comparison to spectroscopic redshift determination for a total of
    6708 maxBCG clusters. The typical photoz error is $\sigma_{z}$ =
    0.02 for the full sample, falling to $\sigma_{z}$ = 0.014 for the
    redshift range 0.07 $\leq z \leq$ 0.3. The increase in photoz
    errors around $z=0.37$ occurs because this is where the 4000 \AA\
    break, the most significant feature in a typical galaxy spectrum,
    passes from the $g$ to $r$ filters. The clusters used in the comparison span the full range of \Ngals of the catalog. \label{zcomp}}
\end{figure}

One of the strengths of this algorithm is that it is a robust
photometric redshift estimator for the clusters: for the 6708 clusters
in the catalog with spectroscopic redshifts available, the dispersion
between the maxBCG estimated redshift and the spectroscopic redshifts
is $\Delta z = 0.018$, as seen in Figure \ref{zcomp}, and is smaller
for the highest richness clusters. These 6708 systems span the range
of \Ngals\ in the catalog. As we do not have spectroscopic redshifts
for all clusters examined, but are confident in relying upon these
estimates, we will henceforth use the term redshift to mean the
estimated photometric redshift determined by maxBCG.

\section{Background Subtraction} \label{method}
Given a set of cluster centers with well-defined three-dimensional
positions, we need to find the galaxies associated with those
clusters. This section explains how we apply background subtraction
techniques to the SDSS data, and how we check our method by
constructing and examining the radial density profile and the
luminosity function.

Generically, any properties of galaxies in clusters can be described
by some population distribution function (PDF) in a multi-dimensional
parameter space. Properties of each galaxy such as luminosity, color,
star formation rate, mass, and distance from the cluster center may be
used as the parameters of a PDF. Examining how the galaxies of a
particular cluster occupy the parameter space is a way to sample the
PDF; with a large enough set of cluster galaxies, we may statistically
determine the PDF quite well. The overall properties of the clusters
(e.g. cluster mass or X-ray temperature) may be used to identify
different sets of clusters, and the PDFs of galaxies in these
different cluster samples compared. In this way, we can explore how
the properties of galaxies in clusters are related to the
characteristics of the host clusters.

Without redshifts for all galaxies, we can only examine the PDF of
galaxies associated with clusters by making an appropriate correction
for the set of field galaxies projected by chance along the line of
sight to the clusters. We assume the presence of a cluster at some
redshift does not affect field galaxies found along the same line of
sight. That is, the PDF of galaxies projected around a cluster center
has two independent components: the distribution of real
cluster galaxies, and the distribution of random background and
foreground galaxies. To determine the projected, azimuthally averaged
PDF of just the galaxies associated with clusters (the PDF$_{C}$), we
examine the PDF of all galaxies projected around cluster centers (the
PDF$_{CF}$) and the PDF of all galaxies projected around a set of
random (field) points on the sky (the PDF$_{F}$). The PDF$_{C}$ is
determined by subtracting: PDF$_{C}$ = PDF$_{CF}$ -
PDF$_{F}$. Although we cannot identify exactly which galaxies make up
a particular cluster, we can very accurately describe the mean
properties of galaxies associated with a set of clusters.

There are a variety of ways in the literature for measuring the
contribution of non-cluster members without having spectroscopic
information. Historically, the population of field galaxies was
estimated from number-flux counts in separate surveys
\citep[e.g.][]{Abe58, Lug86, Col89}, although this method has the
disadvantage of not having the cluster and field samples measured in
the same set of data.  More recently, some authors \citep{Val97,
  Pao01, Got02, Pop05} have measured the background in an annulus
centered on the cluster, in order to ensure that the background
measurement is made using data of similar depth and seeing as the data
in the region of the cluster. Other authors, such as \citet{And04}
estimated the background from a nearby control field, or from the
$logN - logS$ relationship from the same dataset from which the
cluster sample is drawn \citep{lin04}. \citet{Gar99} subtracted
interlopers on the basis of color information, removing ``galaxies
with colors not matching the expected ones at the cluster redshift.''
As galaxy surveys increase in area, it becomes feasible to measure the
background counts directly from the general field, as done by \citet{Gla05}. The SDSS data offer large regions of sky measured to
the same depth and with the same seeing, so we are able to determine
the contribution of field galaxies in the same data as the clusters
without artificially restricting the field measurement to the cluster
neighborhood, or making assumptions about the color or luminosity
distribution of cluster members. By using a set of random points as
the locations around which the field galaxy population is determined,
we measure the characteristics of all galaxies that are associated
with clusters.

\subsection{Application to SDSS data}
The SDSS is an ideal dataset with which to examine the PDF of cluster
galaxies because it provides sky coverage for a large number of
clusters and ample blank sky for measuring the field distribution. The
SDSS data offer a rich parameter space with which to define the
PDF. Properties such as luminosity, color, star formation rate, and
morphology may all be explored. In this work, for $g$-, $r$-, $i$- and
$z$-bands, we construct and examine the \mycpdf: the
density function of cluster galaxies per surface area in a
three-dimensional space of cluster richness \Ngals, projected radius
$r$, and absolute magnitude $M$. The PDF may then be projected onto the
axis of absolute magnitude to show the luminosity function of cluster
members, or onto the axis of projected radius to show the radial
density profile of the cluster.  In this section we describe the
samples of galaxies examined, and as the PDF$_{CF}$ and PDF$_{F}$ are
constructed in the same manner, discuss the construction of a general
\mypdf.

\subsubsection{Cluster and Field Samples}
To measure the PDF$_{CF}$, we examine galaxies projected near the
12830 maxBCG objects found in the redshift range 0.07 $\leq z \leq$
0.3. These systems have richnesses in the range $2 \leq$\ \Ngals\ $\leq 66$. We use all galaxies projected within 2$h^{-1}$ Mpc of the cluster
centers and in the absolute magnitude range -24 $\leq M \leq$
-16. We bin the data in 50 $h^{-1}$ kpc radial x 1 \Ngals\ richness x 0.2
absolute magnitude bins. The number of galaxies in each bin is then
normalized by the physical area observed in each bin. Details of the
calculations are discussed below.

To determine the PDF$_{F}$, we examine galaxies along lines of sight
to randomly chosen field locations. For each cluster, we choose five
positions on the same $\sim$ 200 deg$^{2}$ stripe of sky as the
cluster, with random RA and DEC. These field positions are assigned
the same redshift as the cluster, and labeled with the richness of
that cluster. The resulting set of 64,150 field points are observed
with the same seeing and to the same depth as the clusters, and for
any set of clusters there is a set of field positions with the same
redshift distribution. The same radial and magnitude ranges and bin
widths used for the PDF$_{CF}$ are applied to determine the PDF$_{F}$.
All excess galaxies seen around cluster locations as compared to the
field are identified as cluster galaxies.

\subsubsection{Absolute Magnitudes}\label{absmags}
To calculate absolute magnitudes $M$, apparent magnitudes $m$ must be
corrected for luminosity distance, Galactic dust extinction, and
$K$-corrections as 
\begin{equation}
M = m - 5log_{10}\left(\frac{D_{L}(z)}{10pc}\right) - R -K(z),
\end{equation}

where $D_{L}(z)$ is the luminosity distance for our assumed cosmology;
$R$ is the correction for reddening, computed following \citet{Sch98};
and $K(z)$ is the appropriate $K$-correction.

\begin{figure*}
  \epsscale{1.1} \plottwo{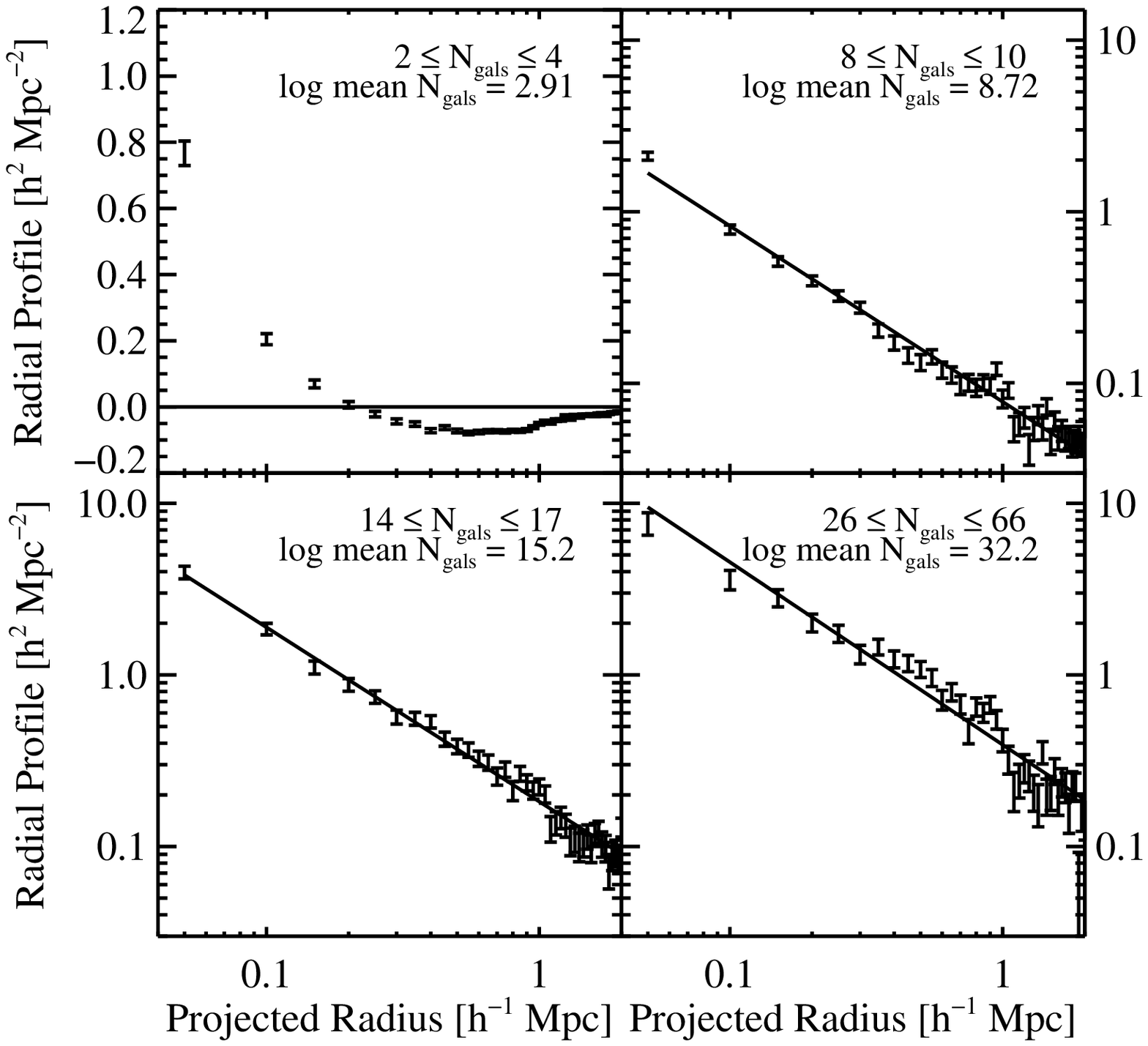}{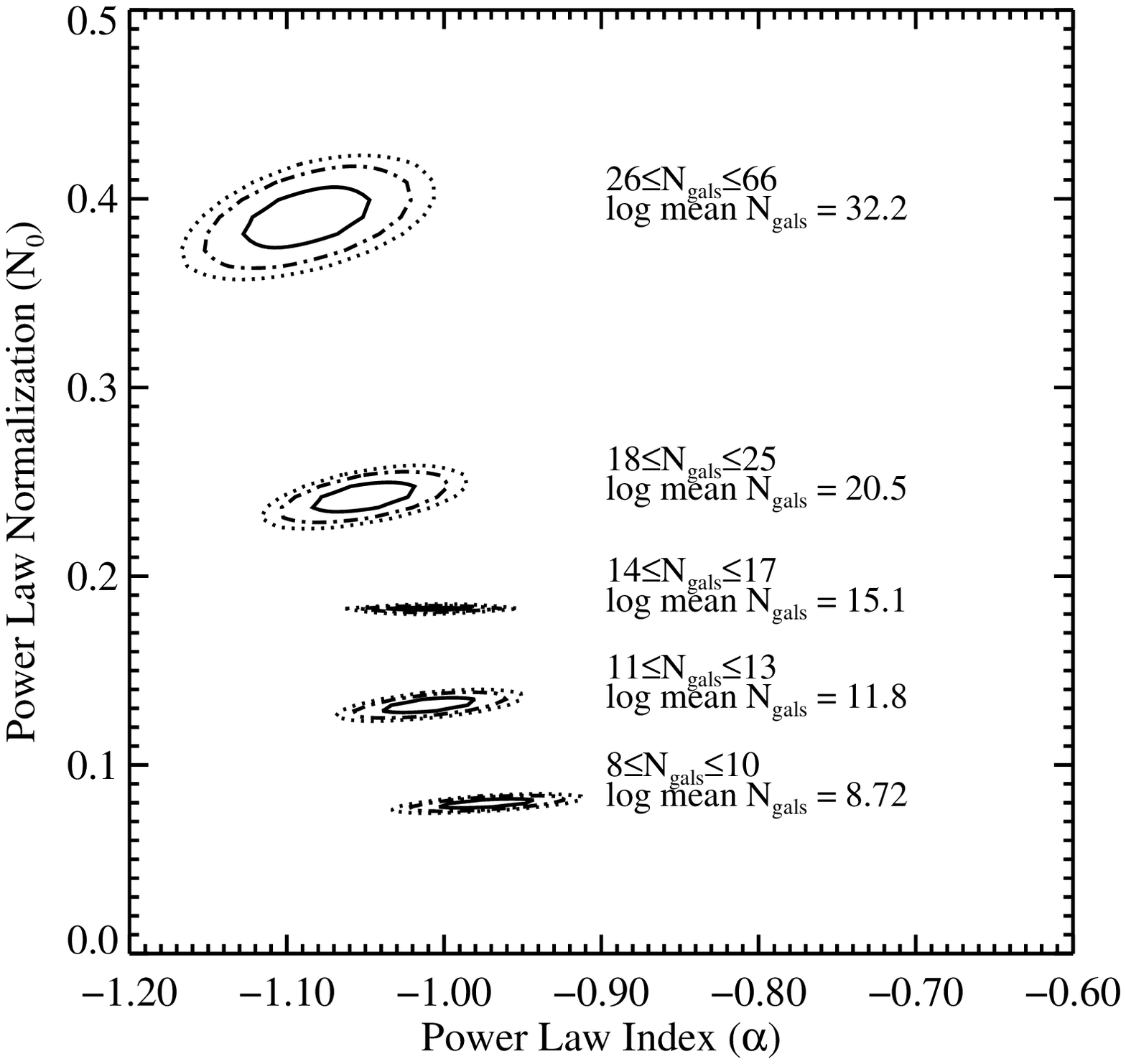} 
\caption[f4a.eps] {
    {\bf Left:} Radial profiles of clusters for selected bins in richness. Shown
    are very poor (upper left), poor (upper right) medium (lower left),
    and high (lower right) richness clusters. The galaxies identified as
    BCGs have been removed. The lowest richness objects tend to be in
    under-dense regions. For
    clusters with \Ngals\ $\ge$ 8, a power law is a reasonable fit to the
    profiles. The best fitting power law is shown.  {\bf Right:} Best-fit
    power law parameters for the radial profiles of clusters of different
    richness. Shown are 1-, 2-, and 3-$\sigma$ $\chi^{2}$ contours of the model parameters for all richness bins with \Ngals\ $\ge$ 8.
    \label{rp3}}
\end{figure*}

$K$-corrections are necessary because galaxy magnitudes in the
observed bandpasses correspond to different rest-frame magnitudes
depending on the redshift of the galaxy. To compare magnitudes of
galaxies at different redshifts, we apply a $K$-correction to convert
all the magnitudes to a fixed set of bandpasses. To do so, we use the
method of \citet{Bla_kcorr}, and following \citet{Bla03}, $K$-correct
all galaxies to $z=0.1$. This redshift is chosen since it is close to
the median redshift of the SDSS spectroscopic sample, and thus
requires the smallest typical corrections.  

Although the true
redshifts of the galaxies are unknown, we apply $K$-corrections and
calculate luminosity distances as though all galaxies projected around
a given point (cluster center or field location) are at the same
redshift as that point. For the galaxies distributed at different
redshifts along the line of sight, the resulting absolute magnitudes
are not correct, but for galaxies actually located at the redshift of
the given position (i.e. those galaxies physically associated with the
cluster) the $K$-correction and $D_{L}(z)$ are appropriate. When the
PDF$_{F}$ is subtracted from the PDF$_{CF}$, the contribution from
galaxies not at the redshift of the cluster is removed, leaving only
the cluster galaxies for which $D_{L}$ and $K(z)$ are correctly
determined.

Our color-dependent $K$-corrections also affect the absolute magnitude
to which the sample is complete and volume-limited at a given
redshift. Galaxies of different colors (therefore with different
amounts of $K$-correction applied) and different apparent magnitudes
can have the same absolute magnitude. Thus $K$-correcting a range of
uncorrected absolute magnitudes $\delta M_{uncorr}$ maps these
magnitudes to the same $M_{kcorr}$. For example, at $z=0.3$, $\delta
M_{uncorr}$\ $\sim$ 0.4 mag; at that redshift a galaxy with apparent
r-band magnitude of 21.0 (the survey limit) corresponds, prior to
$K$-correction, to an absolute magnitude of -19.9. To be complete to
$z=0.3$, we can only use those $M_{kcorr}$\ for which the
corresponding range of $\delta M_{uncorr}$\ does not extend fainter
than -19.9. The end result is that we adopt more conservative
completeness limits to avoid color bias at faint luminosities. We
do not use data in $u$-band because both $K$-corrections and star-galaxy
separation are not as robust in this passband. The resulting absolute
magnitude limits for a complete, volume-limited to \z\ = 0.3 sample of
galaxies in \griz\ are thus -20.2, -19.6, -19.4, and -20.6
respectively.

\subsubsection{Effects of Geometry and Luminosity}\label{areacorr}
We correct for incompleteness both in geometry and in luminosity.
Geometric incompleteness occurs when the search radius around clusters
extends beyond the boundaries of the survey.  For example, at a
redshift of \z\ = 0.07 (the lowest redshift cluster considered here),
the 2$h^{-1}$ Mpc radius aperture is $\sim 0.75$\degr in diameter; some
clusters lie too close to the edge of our 2.5\degr-wide stripe of sky
to have all galaxies within the desired aperture contained on an
observed region of sky.  For each radial bin of each cluster, 
we account for this geometrical incompleteness by calculating
the area that lies on an observed region and weight the galaxy counts 
in each bin accordingly.

Luminosity incompleteness arises because the apparent magnitude limit
of the survey causes a varying range of absolute magnitudes to be
accessible at varying redshifts. For example, to a redshift of 0.3, we
can only see galaxies with $M_{r} < -19.6$ but can examine all
clusters in our catalog, while at \z\ = 0.07, we can see to $M_{r} \sim\ 
-16.5$, but are limited to only a few clusters. To avoid restricting
ourselves to studying only galaxies brighter than the completeness
limit of the full set of clusters (to $z=0.3$), we account for this
luminosity incompleteness. For each magnitude bin, we determine the
number of clusters at redshifts low enough to have galaxies observable
to that limit. The galaxy counts in each bin are weighted accordingly.
The result is that the bright end of the luminosity distribution is
based on galaxies in all clusters in the catalog (at redshifts out to
$z=0.3$), but the faint end is determined from galaxies associated
with lower-redshift clusters only. For our determination of \rN, we
use absolute magnitude limits that ensure luminosity completeness for
the full set of clusters, and push fainter only for examining the
luminosity function of cluster galaxies.

Having calculated the radial and absolute magnitude distributions of
galaxies, we can determine the normalized PDF(\Ngals,r,$M$) per surface
area for galaxies around any set of positions and test our algorithm.

\subsection{Consistency Checks} \label{tests}

In this section we check the background subtraction technique by examining
the radial distribution and the luminosity distribution of galaxies
around field locations, and compare these distributions to those measured around cluster centers.

Note that only a background-subtracted PDF (e.g. the PDF$_{C}$)
contains physically meaningful information; the PDF around any set of
points before subtraction is dominated by galaxies projected by chance
along the same line of sight, which have not been properly
$K$-corrected. We therefore reserve the name `luminosity function' for
the projection of a background-subtracted PDF onto the axis of
absolute magnitude; such a projection for a non-subtracted PDF we
refer to as a `luminosity distribution.' Likewise, we will reserve the
term `radial profile' for background-subtracted PDFs only, and will
refer to the `radial distribution' of galaxies when discussing a
non-subtracted PDF.

A check on the errors recovered by our background subtraction
technique is done by comparing the PDF$_{F}$ with the population density of
galaxies around a set of random points, which are
different random points than those used in constructing the
PDF$_{F}$.  The PDFs measured around two different sets of random points should be statistically identical.

\subsubsection{Radial Number Density Profile} \label{radprof}

A radial number density distribution is constructed by projecting
a \mypdf\ onto the axis of projected radius. Examining the
distribution of galaxies around field locations enables us
to check our correction for incompleteness due to geometry.

When we construct the radial profile of clusters, we first examine the
profile using only galaxies brighter than the completeness limit for
the full set of clusters, and then using all galaxies while accounting
for the varying completeness limit as described above. The results are
statistically the same, but allow for inclusion of fainter galaxies in
the latter case.

The radial
distribution of galaxies around random points is flat, demonstrating
that we are properly calculating the area observed when correcting for
geometrical incompleteness. Comparing the radial distribution
around field locations with that measured around a different set of random points, we see that the distributions are statistically identical at all radii. Any differences are within the error bars, which reflect the Poisson fluctuation of our sample.

\begin{figure}
  \epsscale{0.95} \plotone{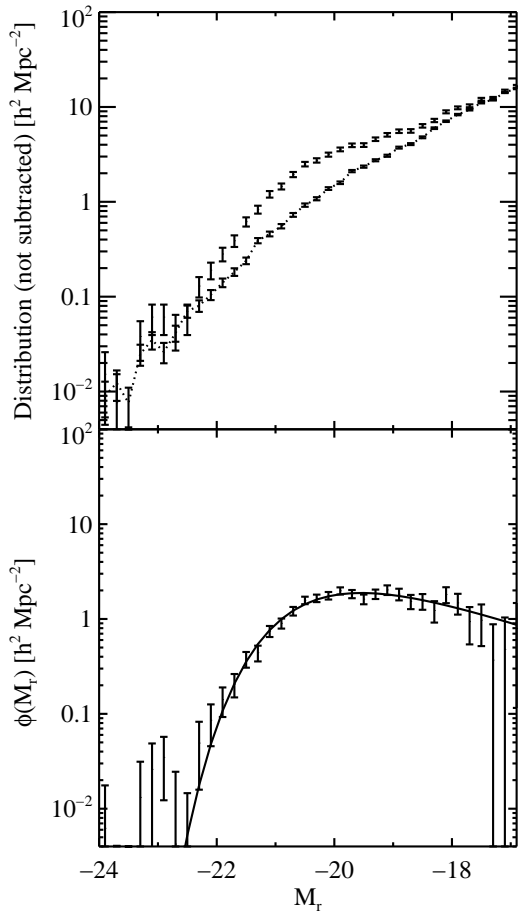}
  \caption[f5.eps] {Luminosity function per unit surface
    area in $r$-band for galaxies in rich clusters (30 $\leq$
    \Ngals\ $\le$ 66), using galaxies within a fixed 1.5$h^{-1}$ Mpc
    aperture. {\bf Top (before subtraction):} luminosity distribution of
    galaxies projected around cluster centers and field points
    (dotted line). Because we $K$-correct all galaxies to the cluster redshift regardless of the true redshift of each galaxy, the luminosity
    distribution around random points does not look like the luminosity
    function of galaxies as measured by e.g. \protect \citet{Bla03}. {\bf Bottom (after subtraction):} luminosity function of
    galaxies associated with these rich clusters. The solid line is the
    best-fitting Schechter function.
\label{lftest_rand}}
\end{figure}

Around cluster centers, however, we find a significant excess of
galaxies compared to the field. This excess varies as a function of
radius and as a function of richness. The left-hand panel of Figure
\ref{rp3} shows the radial profile for selected richness bins
corresponding to very low, low, medium, and high richness clusters,
with the galaxies identified as BCGs removed.  For clusters with
\Ngals\ $\ge$ 5, a power law is an acceptable fit to the profiles. For
sets of clusters of different richness and \Ngals\ $\ge$ 8, the radial
profile is roughly consistent with $\sim r^{-1.1}$ surface density
profile, and thus a $r^{-2.1}$ volume density profile, but with
increasing normalizations for richer clusters, reflecting the larger
size of more massive clusters. For all richness bins above \Ngals\ $\ge$ 8, the $\chi^{2}$ contours of the power
law parameters are plotted as a function of cluster richness in the
right-hand panel of Figure \ref{rp3}. In \S \ref{rpr200} we further
examine the radial profiles, using our measurement of \rN\ to fit a
\citet{nfw97} profile.

\begin{figure*}
  \epsscale{1.} \plottwo{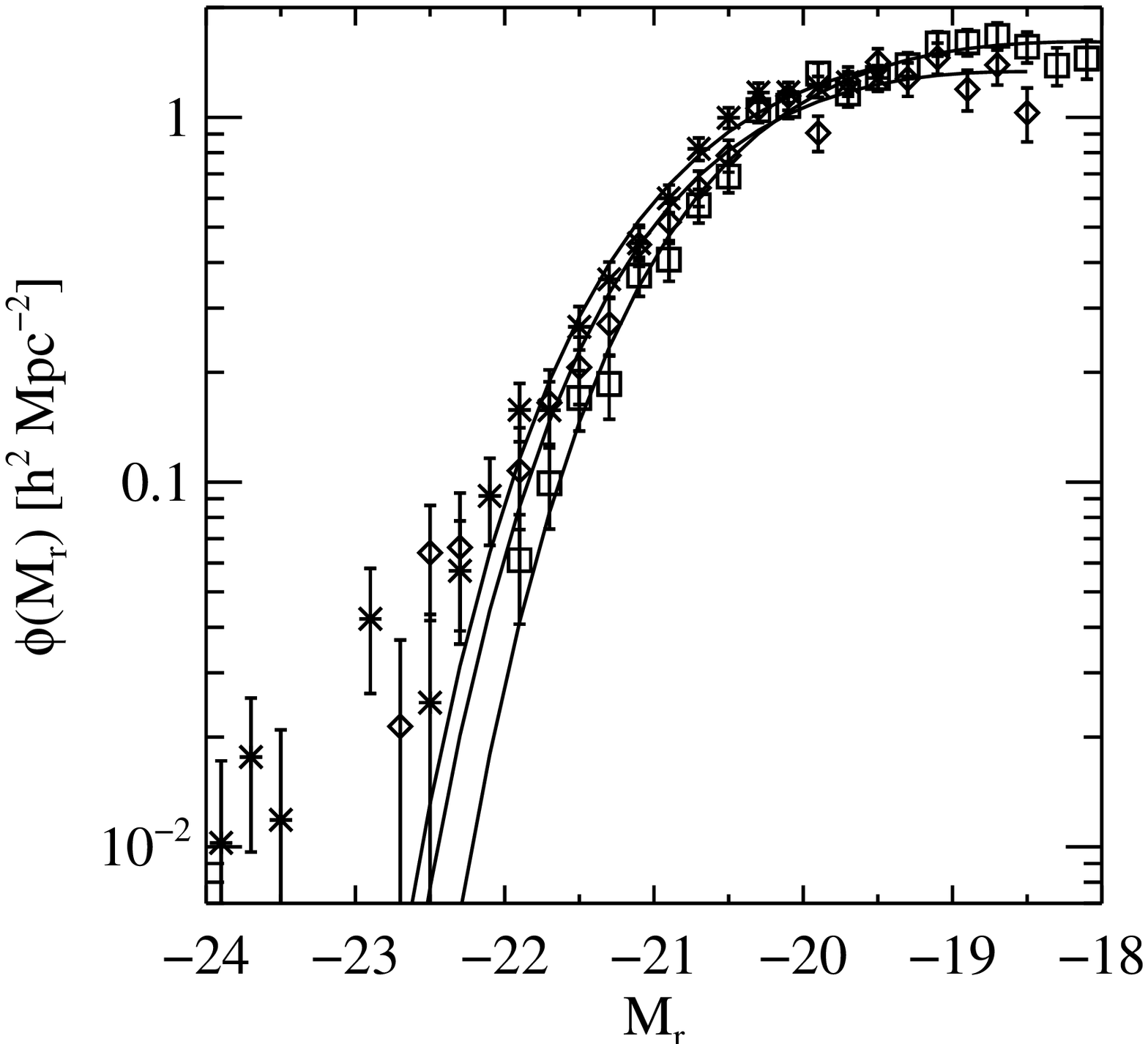}{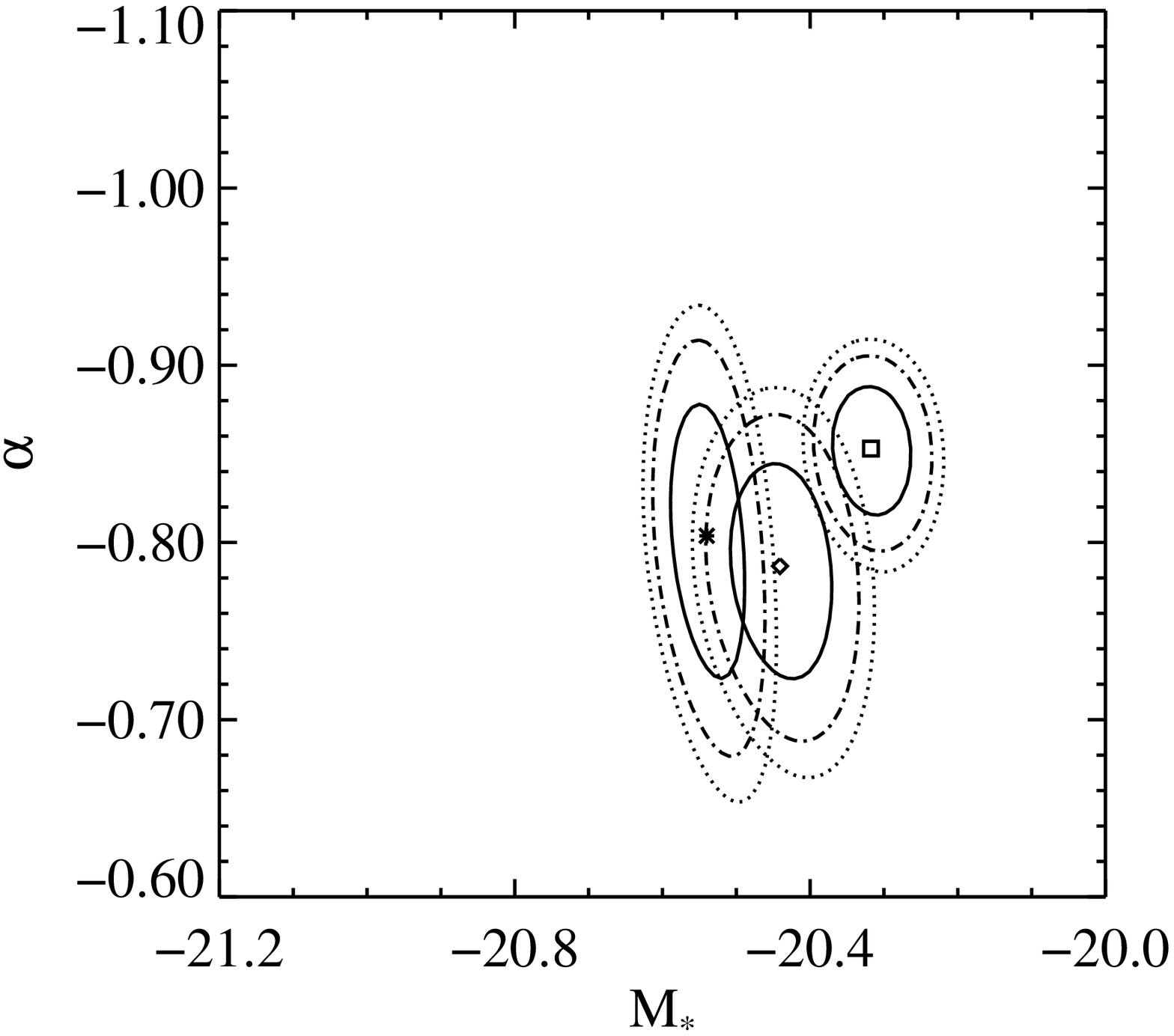} 
\caption[f6a.eps] { {\bf Left:}
    Luminosity function per unit surface area in $r$-band for clusters
    with 18 $\leq$ \Ngals\ $\le$ 66 in three redshift slices. Cluster
    members used are projected within 1.5$h^{-1}$ Mpc of the cluster
    center. The BCGs have been removed. Squares are for clusters in
    0.07 $\le$ z $\le$ 0.15; diamonds for 0.15 $<$ z $\le$ 0.20; stars
    for 0.20 $<$ z $\le$ 0.30. The best-fitting Schechter functions
    are overlaid. {\bf Right:} The 1-, 2-, and 3-$\sigma$ $\chi^{2}$
    contours of the Schechter function parameters.  The shift in the
    characteristic luminosity, \Mstar, reflects the passive evolution
    of the early-type galaxies.
\label{lf3z}}
\end{figure*}

The lowest richness objects (2 $\le$\ \Ngals\ $\le$ 4) tend to be in
under-dense regions.  That is, compared to the distribution of
galaxies around a random point in the universe, these systems of just
a few red galaxies tend to have few nearby neighbors. We do not expect
these very low \Ngals\ objects to be representative of the full
population of low mass halos, as selection effects of the maxBCG
algorithm are significant for these systems. Nonetheless, such a
sample is interesting; we discuss very low \Ngals\ systems in more
detail in \S \ref{disc}.

\subsubsection{Luminosity Function} \label{lumfunc}
The luminosity function of galaxies in clusters is the projection of
the \mycpdf\ onto the absolute magnitude axis. The result is the mean
number of galaxies per cluster per unit surface area as a function of
luminosity and of richness. 

We check our background
subtraction and incompleteness correction by comparing the luminosity
distribution around two different sets of random points, expecting no significant difference. We begin by selecting only locations with low
redshift, so that the luminosity distribution may be examined to faint
magnitudes without weighting for the redshift distribution, and then
also check the distribution with the full sample and appropriate
weighting as discussed in \S \ref{areacorr}. The two samples have identical distributions.

The luminosity distribution contains a statistically significant
excess of galaxies around cluster centers compared to galaxies around
field locations. Figure \ref{lftest_rand}
shows our determination of the luminosity function of galaxies that
are in rich clusters (30 $\le$ \Ngals\ $\le$ 66) and are within
1.5$h^{-1}$ Mpc of the cluster center, both before and after subtraction
of the field (top and bottom figures respectively). In the bottom
figure the solid line plots the best-fitting Schechter function
\citep{Sch76}, of the form

\begin{eqnarray}
\lefteqn{\phi(M)dM=}\nonumber\\
&& 0.4{\rm ln}(10)\phi_{*}10^{-0.4(M-M_{*})(\alpha+1)}e^{-10^{-0.4(M-M_{*})}}dM
\end{eqnarray}

where $\alpha$ is the faint-end slope and \Mstar\ is the turnover
magnitude; we fit the data using the Levenberg-Marquardt $\chi^{2}$
minimization procedure. We recover a luminosity function that is
comparable to that of rich clusters found by \citet{Got02}, who used
different cluster-finding and background subtraction algorithms with
the same sample of SDSS data used in this paper. 

We also examine the LF of cluster galaxies in three redshift slices to
test whether our weighting scheme for the faint end is correct. The
left panel of Figure \ref{lf3z} shows the LF for clusters with 18
$\le$ \Ngals\ $\le$ 66 for 0.07 $\leq z <$ 0.15, 0.15 $\leq z <$ 0.20,
and 0.20 $\leq z <$ 0.30. Galaxies projected within 1.5$h^{-1}$ Mpc
are used; the BCGs are not included. We fit each distribution with a
Schechter function (solid lines). The confidence ellipses for the fit
parameters $\alpha$ and \Mstar\ are plotted in the right panel of the
figure. The only difference we detect between the LFs in different \z\ 
slices is at the bright end, as reflected by the shift of \Mstar\ 
toward fainter magnitudes at lower redshifts. This shift is comparable
to what is expected due to passive evolution of the red sequence
galaxies in the clusters. As our $K$-corrections are non-evolving, we
expect \Mstar\ to be $\sim$ 0.25mag brighter for the highest redshift
bin than for the low \z\ bin due to passive evolution. We note that
these LFs are measured within a fixed physical aperture for systems of
a wide range of richnesses. We present this comparison of luminosity
functions for redshift slices only as a check that we are recovering
sensible LFs. A more detailed investigation of the evolution of the LF
of cluster galaxies will be done in later work.

Since more massive clusters are larger, measuring the LF (or any other
projection of the PDF that varies radially) within a fixed physical
aperture samples different parts of clusters of different
richnesses. For example, as can be seen from the radial profiles in
Figure \ref{rp3}, a 1$h^{-1}$ Mpc radius around a poor group
encompasses the entirety of the group, but only samples the inner
region of a rich cluster. Thus, in order to compare the LF of cluster
galaxies in clusters of different richness, we should examine the LF
of only those galaxies within some aperture that scales with
richness. In addition, the luminosity function of cluster galaxies may
vary with radius. To compare radial trends in clusters of different
richness, we should also use an appropriately scaled aperture. In
\S \ref{r200} we present our calculation of a characteristic
radius of clusters as a function of richness, and will return to
a discussion of the dependence of the luminosity function on richness
and radius in \S \ref{lfr200}.

\section{\rN\ Determination} \label{r200}

\begin{figure}
  \epsscale{1.2} \plotone{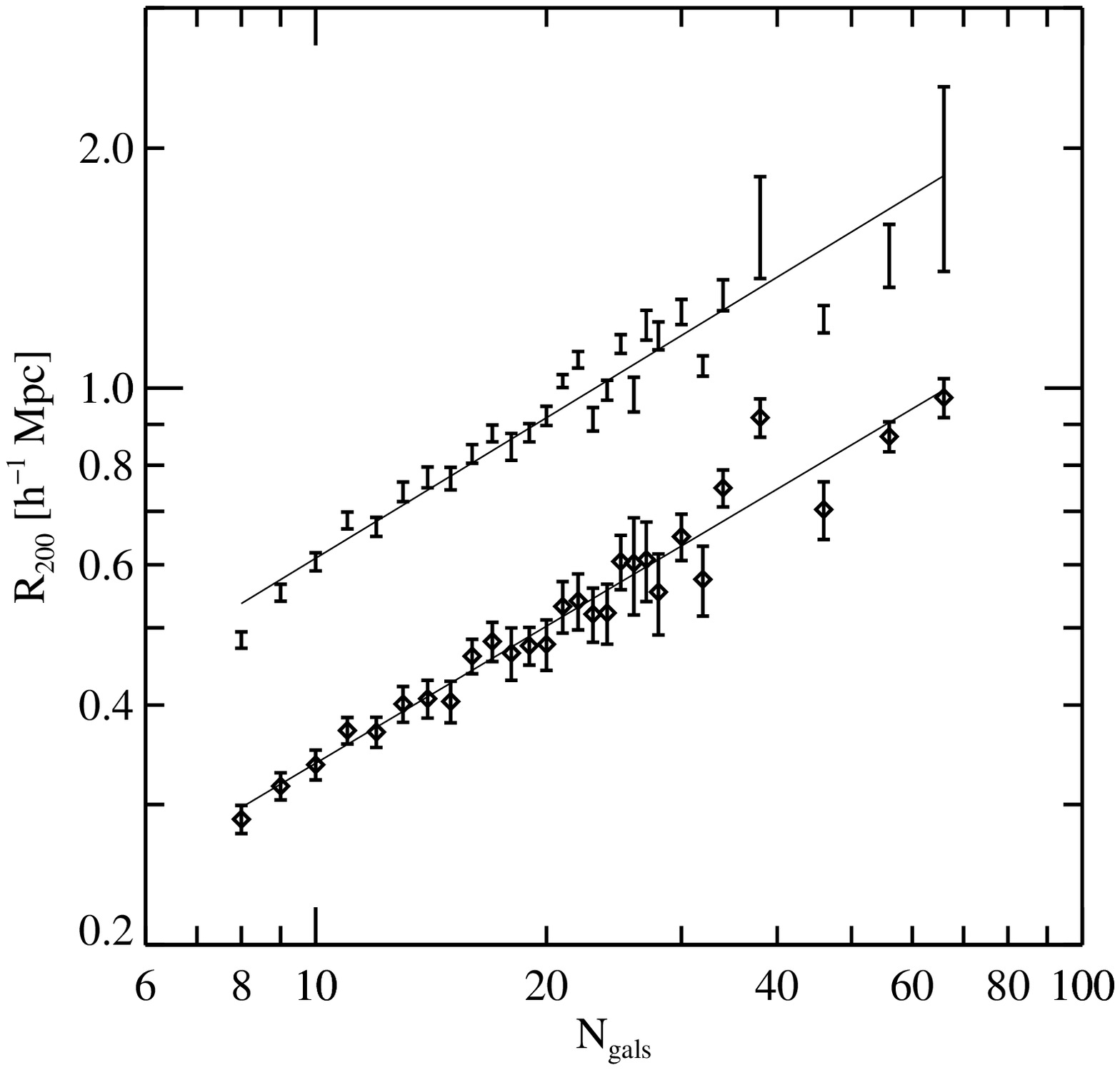} 
  \caption[f7.eps] {Relationship
    of the cluster characteristic radius, \rN, here measured in $r$-band,
    to richness, \Ngals. Diamonds mark the data using the over-density
    threshhold \DNcrit\ $\equiv$ 200 times the critical density; the other data
    points use the over-density threshhold \DNmean\ $\equiv$ 200 times
    the mean density. Using either over-density threshhold results in a
    trend well fit by a power-law with the same scaling. Table
    \ref{r200tab} lists the best-fit values of the
    parameters. \label{r200r}}
\end{figure}

To appropriately compare properties of low- and high-mass objects we
need to understand how the characteristic size of clusters varies with
richness. Motivated by the way \rdelta\ is defined in N-body
simulations, as the threshold radius interior to which the mean mass
density of a cluster is $\Delta$ times the average mass density, we
define an analogous \rN\ using the space density \Ndense\ of galaxies
in clusters compared to the average space density of galaxies. If
galaxies are unbiased with respect to dark matter on all scales, we
would have $R^{\cal N}_{\Delta}$ = \rdelta. Since the bias is
close to unity, we accept $R^{\cal N}_{\Delta}$ as a reasonable
approximation. Following simulations, we use ${\Delta} = 200$ as our
threshold mean over-density of cluster galaxies, which occurs at the
radius \rN. 

Some authors take the average mass density to be the critical density,
while others use the actual mean background density. For the main
result of this paper, the scaling of \rN\ with cluster richness, we
present results both using an over-density threshold of \DNcrit\ $\equiv$\ 200
times the critical density and using \DNmean\ $\equiv$\ 200 times the mean
background density. We find the same scaling using either threshold.
We intend to compare this work with the results using the Hubble
Volume simulations of \citet{Evr02}, who use an over-density threshold
measured with respect to the critical density. Therefore, for
investigations regarding the cluster galaxy population within \rN, we
present results using \DNcrit. Throughout this work, we use the term
\rN\ to mean the radius interior to which the mean number density of
galaxies is \twoOM\ times the mean space density of galaxies, or
equivalently, 200 times the critical density.

\begin{deluxetable}{crrrrr}
\tablecolumns{6} 
\tablewidth{0pc}
\tablecaption{Power Law Fits for \rN(\Ngals) }
\tablehead{
\colhead{} & \multicolumn{2}{c}{$\Delta=200critical$}  & \colhead{} &\multicolumn{2}{c}{$\Delta=200mean$}\\
\cline{2-3} \cline{5-6}\\
\colhead{Band}    &   \colhead{Index}   & \colhead{Normalization}  & \colhead{} &\colhead{Index}   & \colhead{Normalization} }
\startdata 
$g$ & 0.46$\pm$0.03 &  0.17$\pm$0.01  && 0.47$\pm$0.03 &  0.28$\pm$0.02    \\ 
$r$ & 0.57$\pm$0.02 & 0.091$\pm$0.004 && 0.57$\pm$0.01 & 0.159$\pm$0.005  \\ 
$i$ & 0.58$\pm$0.02 & 0.083$\pm$0.004 && 0.60$\pm$0.01 & 0.142$\pm$0.004   \\
$z$ & 0.57$\pm$0.02 & 0.097$\pm$0.004 && 0.58$\pm$0.01 & 0.172$\pm$0.004  \\
\enddata 
\label{r200tab}
\end{deluxetable} 

To determine the mean space density of field galaxies, we use the
\griz\ luminosity functions of \citet{Bla03}, which are
properly normalized to a volume density and are determined with an
SDSS spectroscopic sample of galaxies from the same region of sky. We
integrate these field LFs down to the absolute magnitude limits
applied to our cluster sample (-20.2, -19.6, -19.4, and -20.6 in \griz\ respectively), and take the resulting value to be the average
space density in that passband. We use only these four bands since the
$u$-band $K$-corrections and star-galaxy separation are not as robust as
in these bands.

\begin{figure*}
  \plotone{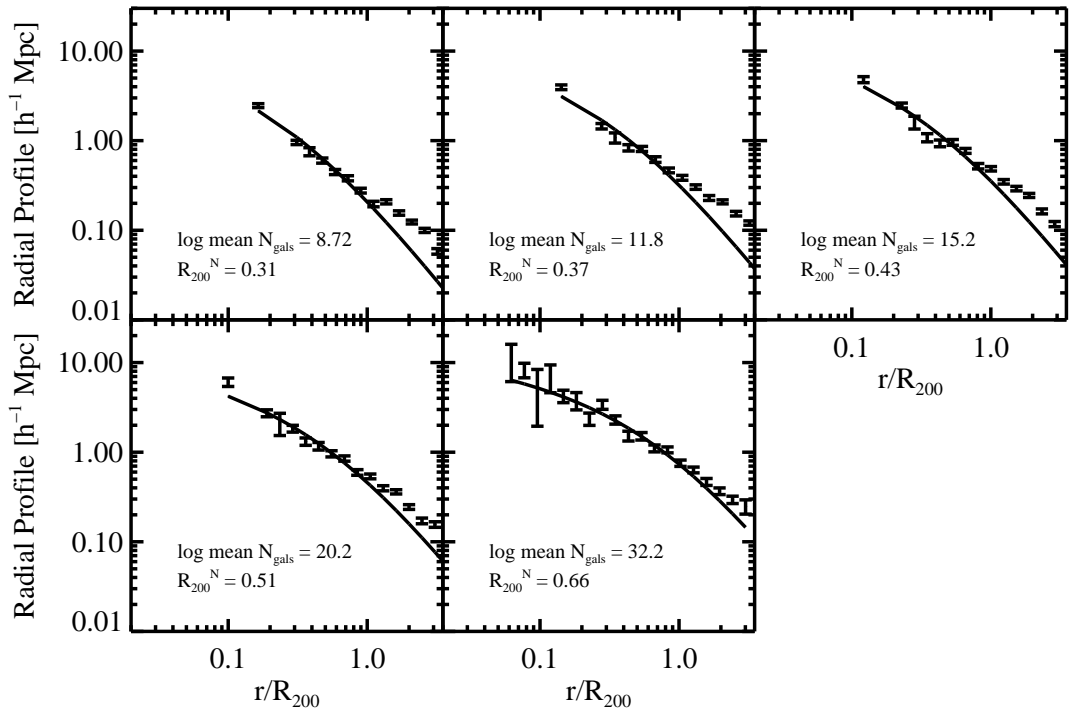} 
\caption[f8.eps]{ Number density
    profiles for satellite galaxies in clusters of different richness.
    The best-fit projected NFW profile is shown, obtained by assuming
    the previously measured relation between \rN\ and \Ngals, and
    fitting the data within \rN\ only. The \rN\ corresponding to the log
    mean \Ngals\ value of the richness bin is listed. The richness bins
    are the same used in examing the radial profiles previously (Figure
    \ref{rp3}). The excess beyond \rN\ is to be expected as the two-halo term becomes important.
    \label{nfw_rp}}
\end{figure*}

To measure the mean space density of cluster members, we use the
\mycpdf\ determined above for each richness. For clusters of a given
richness and in a given bandpass, in each radial bin $r$ we sum over all
bins with radius $\le r$ and with absolute magnitude brighter than the
completeness limit. We assume the galaxies are contained in a sphere of radius $r$ to
calculate the volume density, then divide by the mean space
density of field galaxies to get the fractional excess.  In
actuality, the galaxies are contained in a cylinder of diameter 2$r$.
Thus, although the uncorrelated galaxies are removed from the
measurement, there is an excess of galaxies at radius $r$ due to
the projection.  This excess depends on the shape of the
radial density profile.  For samples that have the same radial
profile, this excess is simply the same multiplicative factor
for all.  For example, if the profile is 1/$r^{2}$, the factor
is $\pi/2$.

We then use the binned mean
over-density vs radius information to find the radius
interior to which the number density of cluster galaxies is
\twoOM\ times greater than the field density. Since the density has roughly a
power law radial dependence, to determine exactly where
$\Delta$\Ndense\ = \twoOM, we fit a line in log-log space in the
region in which the over-density passes through \twoOM. The errors on
this fit include the uncertainty on the density value and the width of
the radial bin, and determine the uncertainty in the \rN\ value. In
this manner we calculate \rN\ for each \Ngals\ bin for which there is
at least one cluster.

The relationship between characteristic radius \rN\ and cluster
richness \Ngals\ is well fit by a power law with index $\sim$ 0.6. The
determination of \rN\ for clusters as a function of richness is the
principle result of this work. Figure \ref{r200r} shows \rN\ measured
with $r$-band data as a function of \Ngals (diamonds), with the best fit
power law plotted. We also plot the relationship between radius and
richness measured using \DNmean. The scaling is the same, with
different normalization. Similar results are obtained for $g$, $i$, and $z$;
the best fit power law parameters with 1-$\sigma$ uncertainties are
listed in Table \ref{r200tab} for all four passbands and both
over-density thresholds.

Under the assumption that the \rN\ we have measured here for clusters
is a good proxy for a mass density based R$_{200}$ for dark matter
halos, what sort of scaling relation would we expect with \Ngals?  A
detailed answer to this question requires understanding the
mass-to-light ratio as a function of both cluster mass and cluster
radius, which is beyond the scope of this paper.  In the simple
scaling arguments below, we assume that these are both constant.  The
radius and mass of a cluster scale as \rtwo\ $\sim M_{200}^{1/3}$, and
the number of galaxies within \rtwo\ is likely to scale as a power law
with cluster mass as N$_{gal\_R_{200}} \sim M_{200}^{\alpha}$.  This
power has been found to be close to unity \citep[e.g.,][]{Kra04,
  lin04, Zeh04, Wec05}.  The \Ngals\ we use here, however, is measured
within a fixed 1$h^{-1}$ Mpc radius aperture; it will typically be
smaller than N$_{gal\_R_{200}}$ for the most massive clusters and
larger than N$_{gal\_R_{200}}$ for smaller groups and clusters, and
roughly consistent with \Ngals\ $\sim$ N$_{gal\_R_{200}}^{\beta}$,
with $\beta<1$.  If the galaxies follow an NFW profile out to max(1
Mpc, \rtwo), with a concentration around 5, one would expect $\beta$
to be in the range $\sim 0.50-0.65$. For maxBCG clusters that have
been found in the simulations of \citet{Wec05}, we find something
similar: \Ngals\ $\sim$ N$_{gal\_R_{200}}^{0.56}$.  Putting this all
together, we have
\begin{equation}
R_{200} \sim  
M_{200}^{1/3} \sim 
N_{gal\_R_{200}}^{1/(3\alpha)} \sim
N_{gals}^{1/(3 \alpha \beta)} \sim 
N_{gals}^{0.6},
\end{equation}
which is in excellent agreement with the scaling relationship that we
find for \rN\ and \Ngals. This comparison suggests that our
observationally determined \rN\ is a reasonably good proxy for \rtwo.
Note that in detail the relation between \Ngals\ measured at a fixed
radius and N$_{gal\_R_{200}}$ is not expected to be a power law over
all halo masses, which implies that the power law relation found
between \rtwo\ and \Ngals\ may break down when measured over a wide
range of halo mass.  This relationship, as well as the detailed
relationship between \rN\ and \rtwo\ as traced by dark matter, will be
explored further in future work, using a larger sample and comparison
to simulations.

\section{Galaxy Density Profiles within \rN} \label{rpr200}
Using our empirically measured \rN, we now examine the radial density
profiles of galaxies in these clusters in greater detail. Simulations
suggest that dark matter halos have mass profiles characterized by a
scale radius $r_{s} \equiv R_{200}/c_{DM}$, where $c_{DM}$ is the
concentration parameter for the dark matter (NFW; \citealt{nfw97}).
Galaxies do not necessarily trace the same detailed distribution as
the dark matter.  In particular, for any given sample of galaxies
chosen with some selection criteria, a range of processes (e.g.,
dynamical friction, tidal stripping, enhanced or suppressed star
formation) may affect the distribution of those galaxies within their
host dark matter halos.  Still, several recent studies have suggested
that the number density profile of galaxies is well described by the
NFW function \citep[e.g.][]{Car97, vdm00, mah04, kat04, lin04}; here we 
fit our radial profiles with the projected NFW profile and examine
its dependence on cluster richness.

We express the number density profile in three dimensions as $n(x) =
n_{0}x^{-1}(1+x)^{-2}$, with normalization $n_{0}$ and $x \equiv
$\cg\ $r/$\rN, where \cg\ is the concentration parameter of the
galaxies. Following \citet{bar96}, we write the projected surface
density NFW as the integral
\begin{equation}
\Sigma(x) = \frac{2n_{0}R^{\cal N}_{200}}{c_{gal}}\int_0^{\pi/2} \cos\theta(\cos\theta + 
\frac{c_{gal}r}{R^{\cal N}_{200}})^{-2} d\theta.
\label{eq:nfw}
\end{equation}

For several bins in richness, we express the radial profile in units
of $r$/\rN, where each cluster has been scaled by the \rN\ appropriate
for its \Ngals\ value, as measured in the previous section and
specified in Table \ref{r200tab}.  We fit the resulting number density
profile within \rN\ with the profile specified in Eq. \ref{eq:nfw};
the results are shown in Figure \ref{nfw_rp}.  The fit does well for
most of the richness bins.  There is some excess outside the virial
radii, which is to be expected as the two-halo term begins to
contribute to the distribution. However, for the low-richness systems,
some of this excess may be due to a misidentification of the cluster
center.  We discuss this issue further in \S \ref{lfr200}.

\begin{figure}
  \plotone{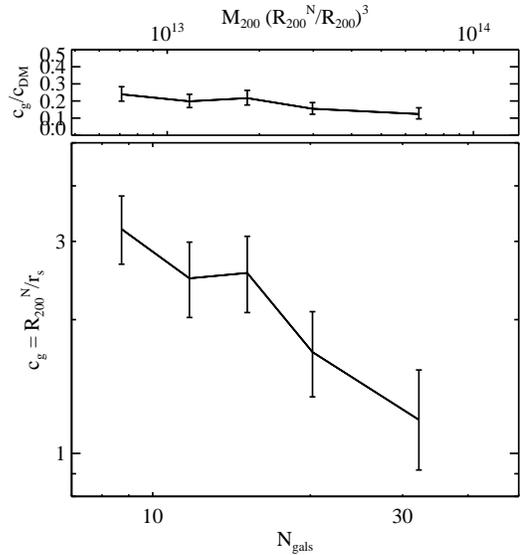} 
\caption[f9.ps]{ Best-fit concentration parameter
    of the galaxy profile, assuming that \rN\ is given by the previously
    determined relation between \rN\ and \Ngals.  The error bars are
    determined from the 1-$\sigma$ region of the $\chi^{2}$ surface.
    The top axis shows $M_{200}$, assuming the same \rN\ to
    \Ngals\ scaling, and under the assumption that \rN\ = \rtwo\ and
    $\Omega_m=0.3.$ 
    \label{nfw_conc}}
\end{figure}

The variation of the concentration parameter with richness is shown in
Figure \ref{nfw_conc}.  Note that we have defined the concentration
parameter for galaxies with respect to the measured \rN\ of the galaxy
profile; if this is not equivalent to the dark matter \rtwo\, then
\cg\ will change accordingly.  In the top of Figure \ref{nfw_conc}, we
show the measured \cg\ divided by the expected dark matter
concentration \cdm, using the model of \citet{Bul01} assuming a
cosmology with $\Omega_m=0.3.$ and $\sigma_8=0.9$.  In order to make
the comparison, we {\em assume} that \rN\ = \rtwo, which may not be
the case, especially for distributions with such different
concentrations.  In particular, preliminary indications from both
simulations and weak lensing measurements indicate that \rN $<$ \rtwo.
Because the theoretical prediction for \cdm$(M)$ gets steeper at higher
masses, in this case \cg$(M)/$\cdm$(M)$ would be closer to a constant
with mass $M$.  

It is clear that the profiles of galaxies in maxBCG clusters have
significantly lower concentrations than the dark matter profiles
measured in CDM N-body simulations.  This finding is in agreement with
previous work \citep[e.g.][]{lin04, Car97, vdm00}, which has found of
\cg\ $\sim 2-4$ for cluster galaxies.  These low values, however,
should be interpreted as indications of how galaxies are distributed
within dark matter halos, and not of the concentration of the dark
matter of these halos. This point was emphasized by \citet{NK04} in
their investigation of a set of hydrodynamic simulations of clusters,
in which they found values of \cg\ $\sim 2-7$ for eight clusters where
the dark matter concentrations were $\sim 6-16$.  In general, for any
population of galaxies in a host halo, the radial distribution is
dependent on the dynamical and star formation histories of the
galaxies once they enter the host halo, and may depend sensitively on
how the population is selected \citep{Die04, Gao04, NK04}.
\citet{Man04} have found that low values of \cg\, quite similar to
what we measure here are required to match the galaxy-galaxy lensing
observations in SDSS.

\section{Cluster Luminosity Functions for $M_{r} < -18$ within \rN} \label{lfr200}
\begin{figure}
  \epsscale{1.1} \plotone{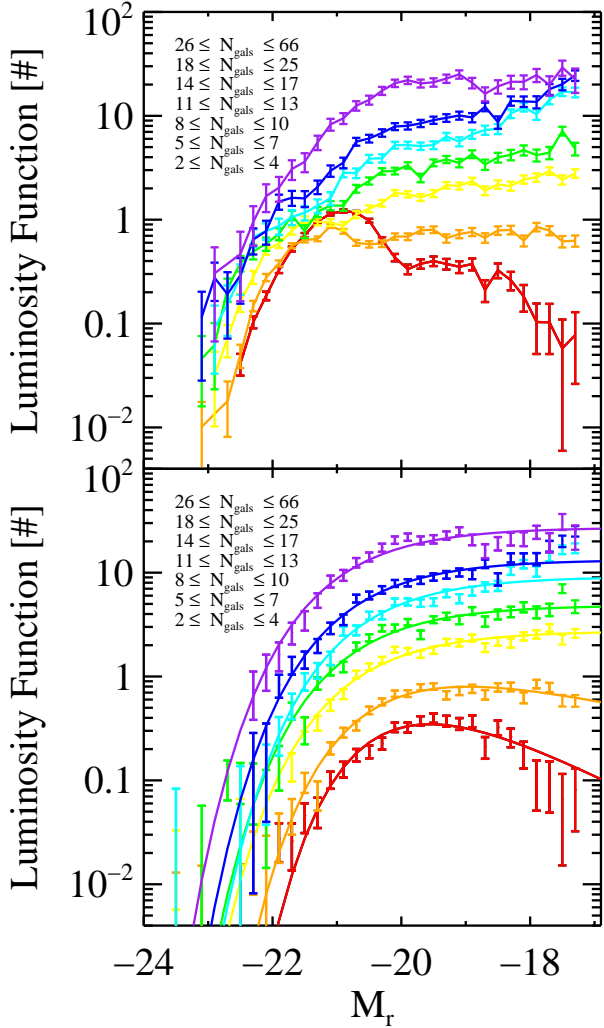} 
\caption[f10_color.eps]{Luminosity
    function within \rN, expressed as the number of galaxies per
    magnitude, for galaxies in clusters of different richnesses.  {\bf
      Top:} LFs of all galaxies within \rN. A Schechter function is
    not a good fit to many of the richness bins, as the effect of the
    BCG is significant.  {\bf Bottom:} LFs for the same sets of
    clusters as above, but with those galaxies identified as BCGs
    removed. Schechter functions fit for $M_{r} < -18$
    are shown; the faint end slope was held fixed for bins with \Ngals\ $\ge$ 8. The parameters and $\chi^{2}$ of the fits are listed in
    Table \ref{Schfits_r200}. \label{lfr200plot} }
\end{figure}

We can use \rN\ to compare commensurate regions within clusters of
different richness. We first measure the luminosity function per unit
surface area of cluster members within the appropriate \rN\ for each
richness, then rescale by the area contained within \rN\ to determine
the actual number of galaxies of each brightness within \rN\ for each
richness. We combine the LFs into six bins of \Ngals. The top panel of
Figure \ref{lfr200plot} shows the LFs of all galaxies within \rN; in
the bottom panel those galaxies identified as BCGs have been removed.

With the BCGs removed, a Schechter function provides a reasonable fit
to the data in all \Ngals\ ranges. \citet{Bla04} have examined the
luminosity function of low luminosity galaxies, finding that surface
brightness selection effects bias the LF to lower values fainter than
$M_{r} \sim$ -18. That is, low luminosity, low surface brightness
galaxies tend to be missed. In addition, \citet{Bla04} find that the
field galaxy LF turns up fainter than $M_{r} \sim$ -18, and that the
shape of the LF over a wide range of magnitudes is best fit by a
double Schechter function. In light of the concerns about missing low
surface brightness galaxies at faint magnitudes, we restrict our fit
to $M_{r} <$ -18. For \Ngals $\ge$ 8, $\alpha = -1$\ provides a reasonable fit to the data in this magnitude range. The parameters of the Schechter function fits, with $\alpha$\ fixed, are listed in Table
\ref{Schfits_r200}. The two lowest-richness bins are not well fit with a faint-end slope of -1; in the Table we list the best-fitting Schechter function parameters for these two bins. Detailed investigation of the differences in LF between different richness samples will be done in future work with a more extensive data set. However, we see that the primary change is the increase in normalization of the LF with increasing cluster richness, that the characteristic luminosity, \Mstar, brightens moderately toward richer clusters, and that the Ngals $<$ 8 systems are different from the richer objects. The mean redshift of the
different richness cluster samples changes by $\Delta$ \"{z} $\sim
0.03$, so we do not expect significant luminosity evolution ($\Delta M_{*} <$ 0.1mag). We note that since these LFs are determined using
cluster members that reside within \rN, the local density is the same
on average for these galaxies, so the richness dependence of the LFs
may not be attributed solely to variations in the local environment.

The sensitivity of the shape of the LF to the cluster richness makes
comparison between different catalogs of clusters difficult. Different
definitions of richness and/or different bins of richness will result
in different measured LFs. In addition, other catalogs typically
present results for the luminosity function using galaxies within a
fixed physical aperture, rather than within an aperture that scales
with mass. Some authors rescale individual cluster LFs by cluster
richness before creating a composite LF, but still examine the LF
within a fixed physical aperture. Nonetheless, we do find
qualitatively similar results to other authors. The luminosity
function we find for rich clusters is similar to the LF of rich
clusters presented in \citet{Got02}, who used the same SDSS data, but
a different cluster-finding algorithm and different method of background
subtraction. Our results are also in agreement with those of
\citet{Pop05} (PVB).  Like PBV, we see that the faint end
of the LF picks up below $M_{r} \sim$ -18, even though we are likely
missing some of these faint galaxies due to surface brightness
selection effects. Our LFs in other bands are also comparable to the
measurements of other authors, who typically find a steeper faint end
in bluer bandpasses. Table 1 of PBV lists the Schechter parameters for
composite cluster LFs retrieved from the literature for a variety of
bandpasses. The LFs in lower richness bins are also comparable with
what has been found by other authors. The $8 \le$ \Ngals\ $\le 10$
groups are comparable to those of \cite{Mar02}, with velocity
dispersions $\sim$ 300 km s$^{-1}$; we find a similar result to theirs
for the luminosity function of these groups when we, like they,
include the BCG. The LFs of very low richness systems (\Ngals\ 
$\le 8$) have a falling faint end slope, and a bright end that
is dominated by the BCGs of these objects. These are the same systems
that are preferentially found in under-dense regions, as shown in \S
\ref{radprof}. We discuss these systems further in \S \ref{disc}.

\begin{deluxetable}{rrrrc} 
\tablecolumns{6} 
\tablewidth{0pc} 
\tablecaption{Schechter Fits for $r$-band Luminosity Functions Within \rN; BCGs Excluded} 
\tablehead{
\colhead{Richness} &   \colhead{\Mstar} & \colhead{$\alpha$} & \colhead{$\phi_{*}$} & \colhead{$\chi^{2}/d.o.f.$}  
}

\startdata
2 $\leq$ \Ngals $\leq$ 4 & -19.95$\pm$0.12 & -0.38$\pm$0.12& 0.94$\pm$0.07 & 1.2\\ 
5 $\leq$ \Ngals $\leq$ 7 & -20.06$\pm$0.08 & -0.55$\pm$0.07& 1.89$\pm$0.12 & 1.2\\ 
8 $\leq$ \Ngals $\leq$ 10 & -20.65$\pm$0.04 & -1.00 (fixed)& 2.93$\pm$0.09 & 1.8\\ 
11 $\leq$ \Ngals $\leq$ 13 & -20.68$\pm$0.04 & -1.00 (fixed)& 5.29$\pm$0.17 & 2.3\\ 
14 $\leq$ \Ngals $\leq$ 17 & -20.58$\pm$0.05 & -1.00 (fixed)& 9.53$\pm$0.33 & 2.0\\ 
18 $\leq$ \Ngals $\leq$ 25 & -20.70$\pm$0.04 & -1.00 (fixed)& 14.1$\pm$0.47 & 1.1\\ 
26 $\leq$ \Ngals $\leq$ 66 & -20.86$\pm$0.05 & -1.00 (fixed)& 29.9$\pm$1.07 & 1.9\\ 
\enddata

\label{Schfits_r200}
\end{deluxetable} 

\begin{figure}
  \epsscale{1.} \plotone{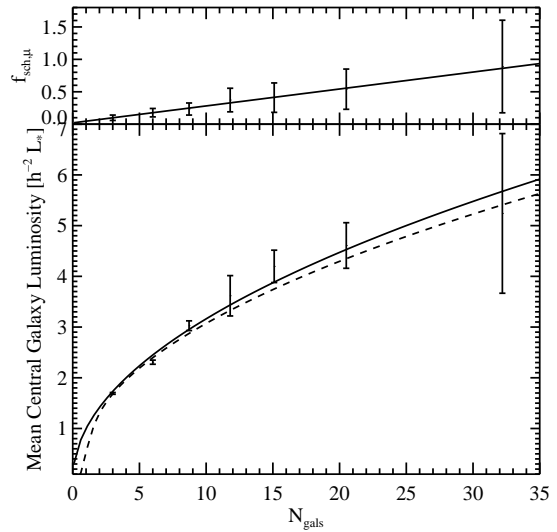} 
\caption[f11.eps]{Parameters related to fitting
  the luminosity functions within \rN, BCGs included, with a model
  comprised of a Gaussian (for the BCGs) plus a Schechter
  function (for the non-BCGs), plotted as a function of the mean log
  \Ngals\ for the ranges used. The LFs used are the same as in Figure
  \ref{lfr200plot}, top panel. {\bf Top:} the ratio of the amplitudes of the
  Schechter and Gaussian functions evaluted at the mean $\mu$\ of the
  Gaussian component. The best-fit line is plotted, and has $f_{Sch,\mu} \sim$ \Ngals/38.3. {\bf Bottom:} The mean luminosity $\mu$\ of the Gaussian fit to the LF of the BCGs, in terms of $L_*$ for SDSS galaxies, as a function of cluster richness. The solid line is $L = $ \Ngals$^{1/2}$; the dashed line is the double power law model of Equation \ref{eq:LNgals}. \label{mungals} }
\end{figure}

With the BCGs included, a Schechter function is not a good description
of the data, except for in the very richest clusters where they are
only a small contribution: the BCG population adds a bright-end
``bump'' to the LF of the other cluster members. The BCGs tend to be
increasingly bright in higher-mass clusters, but also become less
important to the total light with increasing cluster mass.  We fit the
LFs within \rN, BCGs included, with a Gaussian for the BCGs plus a
Schechter function for the non-BCG galaxies. This model is a good fit
to the data for all \Ngals\ ranges. The top panel of Figure
\ref{mungals} shows the relative amplitudes of the Schechter function
and the Gaussian function, $f_{Sch,\mu}$, evaluated at the mean of the
Gaussian component as a function of mean log \Ngals. Over this range
the trend is linear, with $f_{Sch,\mu} \sim$ \Ngals/38.3.  This
scaling is the observed analog for the SDSS to the conditional
baryonic mass function investigated by \citet{Zhe04} (see their Figure
9).

We plot the mean luminosity of the Gaussian as a
function of mean log \Ngals\ in the bottom panel of the Figure. The
trend of brighter BCGs in richer clusters is evident. Over the range we probe, our data are consistent with a single power
law, $L_c \sim$ \Ngals$^{0.5}$.  This result is similar to, but slightly
steeper than, the scaling with mass found by other authors
\citep{lin04BCG, Zhe04, van04}, but note that we are plotting the
scaling as a function of \Ngals\ and not mass.  Our data are all for
multiple-galaxy systems, and as such cannot constrain the very low
mass end of the distribution studied by others.  However, in order to
facilitate comparison with previous work (\citealt{Val04}, and the
compilation of \citet{Coo05}), we fit a double power law function with
several parameters fixed to agree with the low-mass behavior of
\citet{Val04}.  We find that the relationship between mean BCG
luminosity and \Ngals\ is consistent with a model of the form

\begin{equation}
\label{eq:LNgals}
L_c/L_*=\frac{(N_{gals}/N_c)^{4}}{[1+(N_{gals}/N_c)]^{\gamma}},
\end{equation}
where the mean BCG luminosity, $L_c$, is scaled to $L_*$ of the
luminosity function for SDSS galaxies \citep{Bla03}. The best fit
parameters are $N_c = 0.98 \pm 0.04$ and $\gamma = 3.52 \pm 0.02$, but
we stress that these are degenerate with the parameters we have held
fixed; our current data cannot put a strong constraint on the exponent in
the numerator (specifying the scaling for low masses).  We note that
the observation that BCGs are drawn from a different luminosity
function from other cluster members is well known (see, for example,
the review by \citealt{Col03} and references therein). The trend of
increasing central galaxy luminosity with cluster richness has also
been noted in many other observational studies, such as
\citet{San73,San76,Hoe80,Sch83} and \citet{lin04BCG}, and is consistent with predictions from the semi-analytic models of \citet{Ben03} and
\citet{Zhe04} (Z04).

\begin{figure*}
  \epsscale{0.9} \plotone{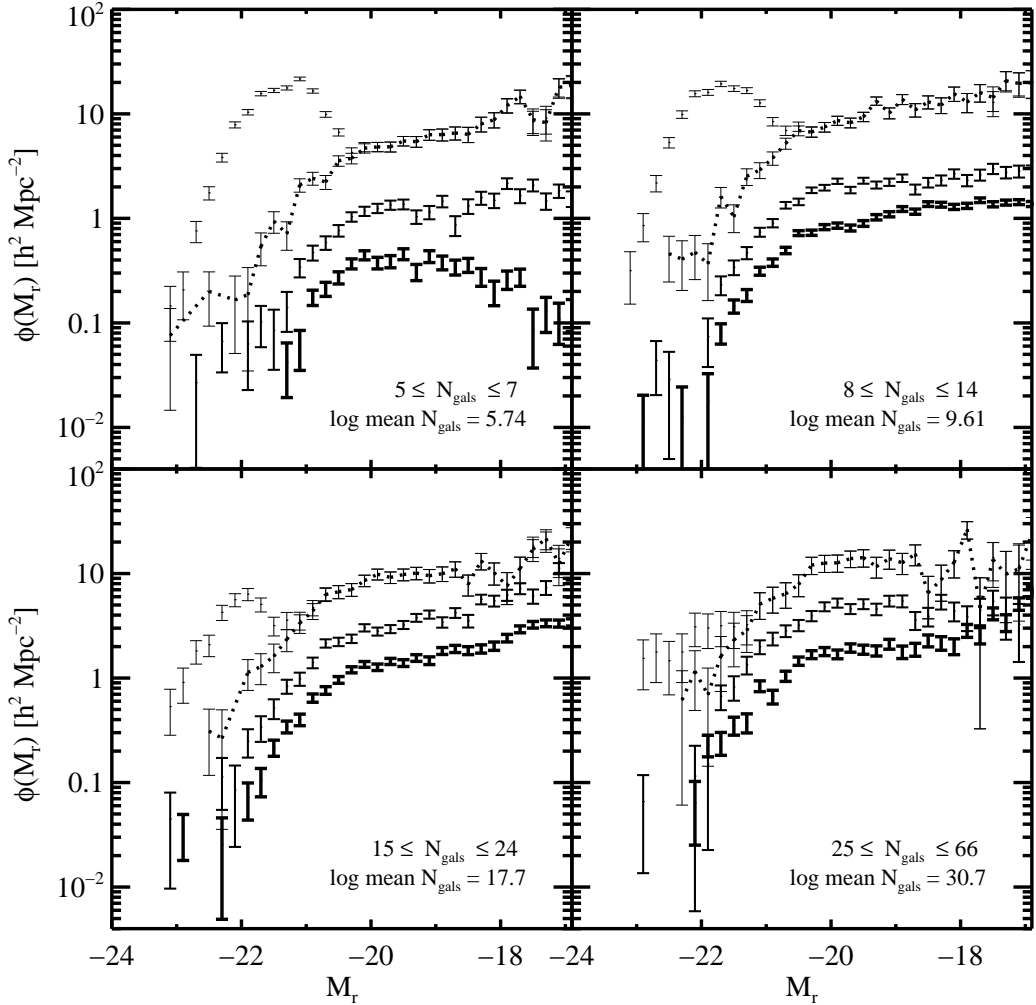} 
\caption[f12.eps]{Luminosity function per unit
    surface area of galaxies in clusters of different richness as a
    function of distance from the cluster center in terms of \rN.
    Light lines are galaxies within $r/$\rN\ $< 0.25$ (the dotted line
    indicates the LF with the BCGs removed); medium-weight lines are
    galaxies in the range $0.25 \leq r/$\rN\ $< 0.75$; the heavy lines
    are galaxies in $0.75 \leq r/$\rN\ $ < 1.75$. For all richnesses,
    the overall normalization of the LF decreases toward larger radii,
    as the density of galaxies drops. Very bright galaxies are found predominantly only in the
    centers of both poor and rich objects. Note that the BCG
    population tends to be brighter and less dominant as cluster
    richness increases. \label{lfradial} }
\end{figure*}

Local density is known to correlate with several galaxy properties,
including luminosity \citep{Bla03}, and so we also expect to see
differences in the LF as a function of radius. Figure \ref{lfradial}
shows the LF of galaxies in three radial bins: $0.0 \leq r/$\rN\ $ <
0.25$\ (light lines), $0.25 \leq r/$\rN\ $ < 0.75$\ (medium lines),
and $0.75 \leq r/$\rN\ $ < 1.75$\ (heavy lines) for galaxies in
clusters in four bins of richness. For the innermost radial bin, we
plot the LF both with and without (dotted lines) the BCGs. For these
LFs, we plot the luminosity function per unit surface area to
explicitly show the change in local density.  For all richnesses, the
overall normalization of the LF decreases toward larger radii, as the
density of galaxies drops. Very bright
galaxies are found predominantly only in the centers of both poor and
rich objects. Except for the lowest richness systems, the faint end slope is roughly similar for all radial bins, although \Mstar\ shifts somewhat fainter towards the outskirts of the clusters. The data are not statistically powerful enough to put strong contraints on these variations; future work with a more extensive data set will allow more detailed investigation.  We have not plotted the data for \Ngals\ $<$ 5 objects,
as just beyond \rN\ of these objects the radial profile becomes
negative, as these objects live in under-dense regions. However, the
LFs of these systems, both with and without BCGs, within 0.25$r$/\rN\ 
is similar to the LF found for 5 $\le$ \Ngals\ $\leq 7$\ groups,
although with a slightly fainter centroid of the BCG population.

Our cluster-finding algorithm identifies the BCG as the galaxy that
maximizes \Lz, a statistic which incorporates luminosity, color, and
the number of fainter neighbors of similar color. There is an
interplay between maximum local density and finding a galaxy with the
right luminosity and color to be a BCG.  Sometimes the maximum total
likelihood selects a galaxy as BCG that the eye would not. The cluster
center is defined to be at the location of the algorithm-defined BCG,
and thus may not be at the true center of the cluster potential if
either the most massive cluster member is not the one with the highest
BCG likelihood, and/or is not at the center of the potential.  Tests
with simulations suggest that the centering of the maxBCG algorithm is
good within $\sim$ 80$h^{-1}$ kpc. For a rich cluster, this amount is
a small fraction of the virial radius, but for a poor group, the
algorithm may misplace the cluster center by a third of the virial
radius. Even if the galaxy identified as the BCG is both the brightest
and the closest to the cluster center, X-ray observations suggest that
a BCG can reside as far as $\sim$ 70$h^{-1}$ kpc from the center of
the X-ray emission \citep{Laz98, lin04BCG}.

The inner region of the lowest richness objects may be biased due to
failing to correctly identify the galaxy most closely located at the
center of the cluster, and therefore incorrectly positioning the
center of the cluster. For the radial profiles, a higher inner radial
bin due to incorrect removal of the BCG of the lower-richness objects
will skew the measured concentration parameter to higher values for
those objects. We also see the presence of BCG-like galaxies in the
very inner regions of the lowest-richness clusters in the luminosity
function: examining the region within 50$h^{-1}$ kpc of the centers of
poor clusters shows a luminosity function with a small ``BCG bump'' at
the bright end, even when the maxBCG-identified BCGs are removed.
Further investigation is needed to show whether these galaxies are the
true BCGs of these groups, or whether the groups host a population of
BCG-like galaxies in addition to the BCG.

\section{Discussion and Conclusions}\label{disc}
By using statistical background subtraction, we have observationally
determined the changes in the radial distribution of cluster galaxies
as a function of cluster richness, and used that information to
calculate a characteristic radius for clusters of each richness. This
model-independent radius, \rN, is based on the number density of
galaxies analogously to the way \rtwo\ in use in simulations is
based on mass density. We find that \rN\ exhibits a power law scaling
relationship with cluster richness. This result is in good
agreement with the expected scaling from simulations.

Our result is also in general agreement with results obtained by other
groups using different methods for measuring richness and \rtwo.
\citet{Gir95} examined spectroscopically confirmed members of 90 rich
clusters, and determined virial and core radii by fitting the observed
galaxy distribution with an isothermal sphere. They find virial radii
of 0.5 - 1 $h^{-1}$\ Mpc, the same range we find for \rN\ for rich
clusters. \citet{Yee03} examined \rtwo\ as a function of cluster
richness $B_{gc}$\ (the amplitude of the galaxy cluster center
correlation function measured for each cluster, scaled by a luminosity
function and spatial distribution) for 16 rich clusters with
spectroscopically confirmed members. They measured \rtwo\ by applying
a singular isothermal sphere model to the velocity dispersion data
\citep{Car97_r200}, and find \rtwo\ $\sim B_{gc}^{0.54 \pm 0.18}$.
This scaling matches well with what we find, although with greater
scatter than our result. With our techniques, we have been able to use
photometric data to confirm this relationship for rich clusters and
extend the relationship to much less rich systems.

To compare equivalent regions of clusters of different richness, \rN\ 
may be used as an aperture within which to compare the properties of
cluster members. The space density distribution of galaxies within
\rN\ is well described by an NFW profile, and the derived
concentration parameter varies with cluster richness. We examined the
population distribution function of cluster galaxies to determine how
the luminosity function of cluster members changes both radially and
with cluster richness, using our determination of \rN\ to compare
clusters in a wide range of richnesses.  The radial variation of the
luminosity function of cluster galaxies is similar in clusters of all
richnesses, but does depend on the cluster richness. It is important
to note that we can still detect a signal from the clusters at
2$h^{-1}$ Mpc from the cluster, even for poor groups. That there is
still a significant over-density at these distances suggests caution
when measuring the contribution of the background in an annulus
centered on the cluster.

We find that the central galaxies of clusters are distinguishable from
the rest of the cluster galaxy population, as has been noted by others
in both observational studies and in theoretical models. The BCG
population is clearly evidenced in the luminosity function within \rN,
and more dramatically in the LF of the central region of clusters, as
a ``bump'' at bright magnitudes that rises above the LF of the other
cluster members. As the richness of the cluster increases, the BCG
population becomes brighter but contributes less to the overall
cluster light, in agreement with the results of $K$-band observations
of BCGs in 93 X-ray selected systems studied by \citet{lin04BCG}. We
find that the luminosity function of low- and intermediate-richness
systems cannot be well described by a Schechter function when the BCGs
are included; this function provides an acceptable fit only for the richer
systems where the fractional contribution of the central galaxy to the
LF is small.  The total cluster luminosity function within \rN\ is
well modeled by the sum of a Gaussian for the central galaxy and a
Schechter function for the satellites over the whole range of \Ngals.
The trends of mean BCG luminosity and fractional contribution of the
BCGs are in good agreement with the models of \citet{Zhe04}. We find that the mean BCG luminosity scales with mean log \Ngals\ as \Ngals$^{1/2}$, a similar though slightly steeper result to what is seen in other studies.

The low-richness objects identified with our cluster finding algorithm
have properties that are quite different from rich clusters. Systems
with \Ngals\ $<$\ 5 and \Ngals\ $>$\ 7 have clearly different radial
profiles and luminosity functions. The transition region $5 \leq $\ 
\Ngals\ $\leq 7$\ has intermediate properties. One possible
contribution to this difference is that the lowest-\Ngals\ objects are
also the most likely to suffer from the effects of misidentifying the
cluster center and/or the the brightest cluster member. In addition,
the \Ngals\ $<$ 8 objects have a strong selection function: maxBCG is
finding systems where specifically only a few red galaxies are within
1$h^{-1}$ Mpc of the galaxy identified as the BCG. It is interesting
that demanding so few galaxies of that color be in the neighborhood
has the effect of finding objects that have very concentrated radial
profiles, are located in isolated regions, and are dominated by bright galaxies.

A possible explanation for these very low \Ngals\ systems is that at
least some of them are fossil groups. Observationally, a fossil group
has the X-ray luminosity of a group or poor cluster, but in the
optical, only a highly luminous early-type galaxy without bright
neighbors is observed. The first such object was detected by
\citet{Pon94}; subsequent study has shown that there are a few other
galaxies in this group, but all are significantly fainter ($\Delta \rm
m \ge 2.5mag$) than the first-ranked galaxy \citep{Jon00}. Other
fossil groups have been identified as well
\citep{Mul99,Vik99,Rom00,Mat01}. Such systems are thought to be the
end-product of groups in which most of the galaxies have merged,
producing a highly luminous central galaxy mostly alone in a halo of
hot X-ray gas. This scenario is supported by the observations that the
central galaxies emit a very high fraction of the total optical light
of the group, and that there is a dearth of $\rm L_{*}$\ galaxies in
the central region of these systems \citep{Jon03}.

At fixed halo mass, there is a theoretical expectation that there
should be a correlation between halo formation time and the total
number of galaxies above a given luminosity threshold \citep{Zent05}.
This correlation may select a special population of of clusters at low
\Ngals.  We speculate that in high mass halos, all systems have bright
red galaxies regardless of their formation time, and are equally well
found by the maxBCG algorithm, while in smaller, group-mass halos the
requirement to select red galaxies may select only the earliest
forming halos.  These low \Ngals\ systems would be expected to be
early-forming halos with higher than average masses for their \Ngals,
whose outer satellites have already merged with the central object.
This theory is consistent with the LF shape and the extremely steep
radial profiles seen here, as well as with the higher masses that are
indicated by cluster-mass correlation function examined by
\citet{she05}.

Many of these systems may also be classified as compact groups (CGs),
such as those studied by \citealt{Hic82,Hic93} and others. Compact
groups have been identified in SDSS data by \citet{Lee04}. They
investigate the local environment of CGs, and find that while on
average the number density of surrounding galaxies is comparable to
the local environment around field galaxies, there is considerable
scatter to both more and less dense environments (see their Figure 9).
By demanding few other bright, red galaxies nearby, the maxBCG
algorithm may be preferentially selecting fossil groups, which would
also look like those CGs in environments with a low number density of
neighbors. Further study is needed to understand this interesting set
of systems. More investigation can be done using SDSS data, but it
also would be interesting to look for an X-ray signal from a stacked
set of these low-richness objects.

We draw attention to the range of luminosity functions for galaxies
within \rN\ of clusters of different richness, and also to the
differences in the LF seen radially. Such variation makes it difficult
to compare between different cluster catalogs, where different proxies
for mass and aperture are in use. The LF of clusters in a mass range
are different when different fractions of \rN\ are sampled, and
different again when a fixed aperture is used for each cluster.
Nonetheless, we do see similar results to those of other authors,
including an upturn at the faint end for rich clusters. This effect
comes primarily from galaxies located in the outer regions of the
clusters.

We are currently working to compare our results with simulations to
show how our space density-based \rN\ relates to the mass
density-based \rtwo\ commonly used in N-body models. Also, we will
examine how our richness parameter, \Ngals, relates to mass $M_{200}$
from the simulations. Further work will be done to use the rich SDSS
dataset to explore the distribution of many properties of cluster
galaxies in addition to those considered here.

Using a small subset of the SDSS data is sufficient to determine the
scaling of \rN\ with cluster richness, and to robustly measure the
luminosity functions and radial profiles in small \Ngals\ bins. Using
the full SDSS dataset will allow us to study $\sim$ 500 of the richest
clusters and over 25,000 groups, and will allow cluster detection to
higher redshifts. We will check for evolution in our scaling
relationship by examining the \rN -\Ngals\ trend in different redshift
samples.  Preliminary results indicate that the scaling is the same,
but with higher normalization for lower redshift objects. To compare
the properties of any stacked set of clusters, the relationship of
\rN\ to \Ngals\ is essential. We can now compare different mass
estimates, including those measured from total luminosity, velocity
dispersion, and from lensing information. We will also use lensing
studies to compare the profiles of the luminous and dark components of
the clusters, more closely investigate the possible differences in
concentration of the two profiles, and examine the bias between
luminous and dark matter in these dense environments. Extending this
work to include color and morphology indicators, we will probe many
characteristics of galaxies in clusters, exploring the history of
galaxy formation as a function of environment, and studying the BCG
population.

For large surveys it is typically not practical to
determine cluster mass or size using spectroscopy. Without assuming a
model for the distribution of galaxies, we provide a way to determine
cluster size from photometric data alone for $z \le$ 0.3.  With
knowledge of how the size-mass scaling evolves and a better
understanding of the scaling of \rN\ with total mass, our method
provides a feasible way to measure the characteristic radii and masses
of clusters that will be found in future large, high-redshift surveys.

\acknowledgments
Funding for the creation and distribution of the SDSS Archive has been
provided by the Alfred P. Sloan Foundation, the Participating
Institutions, the National Aeronautics and Space Administration, the
National Science Foundation, the U.S. Department of Energy, the
Japanese Monbukagakusho, and the Max Planck Society. The SDSS Web site
is http://www.sdss.org/. 

The SDSS is managed by the Astrophysical Research Consortium (ARC) for
the Participating Institutions. The Participating Institutions are The
University of Chicago, Fermilab, the Institute for Advanced Study, the
Japan Participation Group, The Johns Hopkins University, the Korean
Scientist Group, Los Alamos National Laboratory, the
Max-Planck-Institute for Astronomy (MPIA), the Max-Planck-Institute
for Astrophysics (MPA), New Mexico State University, University of
Pittsburgh, University of Portsmouth, Princeton University, the United States Naval Observatory,
and the University of Washington. 

SMH was supported by a General Electric Faculty for the Future
Fellowship and by the Michigan Space Grant Consortium. TAM
acknowledges support from PECASE grant AST 9708232 and NSF grant AST
0206277. RHW was supported by NASA through Hubble Fellowship
HF-01168.01-A awarded by Space Telescope Science Institute.  We thank
Gus Evrard, Andrey Kravtsov, Martin White, and Andrew Zentner for
helpful discussions and suggestions.

\end{document}